\documentclass[letterpaper, 10 pt, journal]{ieeeconf} 

\pdfoutput=1

\IEEEoverridecommandlockouts                              % This command is only
\usepackage{cite}
\usepackage{amsmath}
\usepackage{amssymb}
\usepackage{pifont}
\usepackage{array}
\usepackage{url}
\usepackage{hyperref}
\usepackage[font=singlespacing]{caption}

\newenvironment{tightItemize}{
\begin{itemize}
        \setlength{\itemsep}{1pt}
        \setlength{\parskip}{0pt}
        \setlength{\parsep}{0pt}
}{\end{itemize}
}

\newenvironment{tightEnumerate}{
\begin{enumerate}
        \setlength{\itemsep}{1pt}
        \setlength{\parskip}{0pt}
        \setlength{\parsep}{0pt}
}{\end{enumerate}
}

\title{\LARGE \bf
Data-Parallel Hashing Techniques for GPU Architectures
}

\author{Brenton Lessley% <-this % stops a space
\\
\textit{Computer and Information Science \\
        University of Oregon \\
        {\tt\small blessley@cs.uoregon.edu}}%
}

\begin{document}

\maketitle
\thispagestyle{empty}
\pagestyle{empty}

%%%%%%%%%%%%%%%%%%%%%%%%%%%%%%%%%%%%%%%%%%%%%%%%%%%%%%%%%%%%%%%%%%%%%%%%%%%%%%%%
\begin{abstract}
Hash tables are one of the most fundamental data structures for effectively storing and accessing sparse data, with widespread usage in domains ranging from computer graphics to machine learning. This study surveys the state-of-the-art research on data-parallel hashing techniques for emerging massively-parallel, many-core GPU architectures. Key factors affecting the performance of different hashing schemes are discovered and used to suggest best practices and pinpoint areas for further research.

\end{abstract}

\section{Introduction}

The problem of searching for elements in a set is a well-studied algorithm in computer science.
Canonical methods for this task are primarily based on sorting, spatial partitioning, and hashing~\cite{Knuth:1998}.
In searching via hashing, an indexable hash table data structure is used for efficient random access and storage of sparse data, enabling fast lookups on average.
For many years, numerous theoretical and practical hashing approaches have been introduced and applied to problems in areas such as computer graphics, database processing, machine learning, and scientific visualization, to name a few~\cite{Ullman:1972,Maurer:1975,Karlin:1986,Czech:1997,Knuth:1998,Wang:2014,Wang:2016}.
With the emergence of multi-processor CPU systems and thread-based programming, significant research was focused on the design of concurrent, lock-free hashing techniques for single-node, CPU shared-memory~\cite{Greenwald:2002,Michael:2002,Peierls:2005,Shalev:2006,Goodman:2010}. Moreover, studies began to investigate external-memory (off-chip) and multi-node, distributed-memory parallel techniques that could accommodate the oncoming shift towards large-scale data processing~\cite{Botelho:2007,Cheng:2014}.
These methods, however, do not demonstrate node-level scalability for the massive number of concurrent threads and parallelism offered by current and emerging many-core architectures, particularly \textit{graphical processing units} (GPUs). GPUs are specifically designed for data-parallel computation, in which the same operation is performed on different data elements in parallel.
%GPU-based hashing research acknowledges these needs for exposing more parallelism by suggesting lock-free shared-memory, data-parallel techniques that trade-off different performance aspects, while using hardware atomics that are generally less-complicated and more fine-grained than those in predecessor CPU studies.
%Moreover, very recent work has introduced atomics-free data-parallel hashing for certain use cases, leading a transition into platform-portable hashing for emerging many-core architectures.

CPU-based hashing designs face several notable challenges when ported to GPU architectures:
\begin{tightItemize}
\item \textit{Sufficient parallelism}: Extra instruction- and thread-level parallelism must be exploited to cover GPU global memory latencies and utilize the thousands of smaller GPU compute cores. Data-parallel design is key to exposing this necessary parallel throughput. 
\item \textit{Memory accesses}: Traditional pointer-based hash tables induce many random memory accesses that may not be aligned within the same cache line, leading to multiple global memory loads that limit throughput on the GPU.
\item \textit{Control flow}: Lock-free hash tables that can be both queried and updated induce heavy thread contention for atomic read-write memory accesses. This effectively serializes the control flow of threads and limits the thread-level parallelism on the GPU.
\item \textit{Limited memory}: CPU-based hashing leverages large on-chip caching and shared memory to support random-access memory requests quickly. On the GPU, this fast memory is limited in size and can result in more cache misses and expensive global memory loads. 
\end{tightItemize}
In this study, we survey the state-of-the-art data-parallel hashing techniques that specifically address the above-mentioned challenges in order to meet the requirements of emerging massively-parallel, many-core GPU architectures. These hashing techniques can be broadly categorized into four groups: open-addressing, perfect hashing, spatial hashing, and separate chaining. Each technique is distinguished by the manner in which it resolves collisions during the hashing procedure.

The remainder of this survey is organized as follows. Section~\ref{background} reviews the necessary background material to motivate GPU-based data-parallel hashing. Section~\ref{hashing-techniques} surveys the four categories of hashing techniques in detail, with some categories consisting of multiple sub-techniques. Section~\ref{applications} categorizes and summarizes real-world applications of these hashing techniques at a high-level.  Section~\ref{analysis} synthesizes and presents the findings of this survey in terms of best practices and opportunities for further research. Section~\ref{conclusion} concludes the work.
\section{Background}
\label{background}
The following section reviews concepts that are related to GPU-based data-parallel hashing.

\subsection{Scalable Parallelism}

Lamport~\cite{Lamport:1978} defines \textit{concurrency} as the decomposition of a process into independently-executing events (subprograms or instructions) that do not causally affect each other.
\textit{Parallelism} occurs when these events are all executed at the same time and perform roughly the same work. According to Amdahl~\cite{Amdahl:1967}, a program contains both non-parallelizable, or serial, work and parallelizable work. Given $P$ processors (e.g., hardware cores or threads) available to perform parallelizable work, \textit{Amdahl's Law} defines the \textit{speedup} $S_P$ of a program as $S_P \leq T_1 / T_P$, where $T_1$ and $T_P$ are the times to complete the program with a single processor and $P$ processors, respectively. As $P\to \infty$, $S_{\infty} \leq \frac{1}{f}$, where $f$ is the fraction of serial work in the program. So, the speedup, or \textit{scalability}, of a program is limited by its inherent serial work, as the number of processors increases. Ideally, a linear speedup is desired, such that $P$ processors achieve a speedup of $P$; a speedup proportional to $P$ is said to be scalable. 

Often a programmer writes and executes a program without explicit design for parallelism, assuming that the underlying hardware and compiler will automatically deliver a speedup via greater processor cores and transistors, instruction pipelining, vectorization, memory caching, etc~\cite{Jeffers:2015}. While these automatic improvements may benefit \textit{perfectly parallelizable work}, they are not guaranteed to address \textit{imperfectly parallelizable} work that contains data dependencies, synchronization, high latency cache misses, etc~\cite{McCool:2012}. To make this work perfectly parallelizable, the program must be refactored, or redesigned, to expose more \textit{explicit} parallelism that can increase the speedup ($S_P$). Brent~\cite{Brent:1974} shows that this explicit parallelism should first seek to minimize the \textit{span} of the program, which is the longest chain of tasks that must be executed sequentially in order. Defining $T_1$ as the total serial work and $T_\infty$ as the span, \textit{Brent's Lemma} relates the work and span as $T_P \leq (T_1 - T_{\infty}) / P + T_{\infty}$. This lemma reveals that the perfectly parallelizable work $T_1 - T_{\infty}$ is scalable with $P$, while the imperfectly parallelizable span takes time $T_\infty$ regardless of $P$ and is the limiting factor of the scalability of $T_P$. 

A common factor affecting imperfectly parallelizable work and scalability is memory dependencies between parallel (or concurrent) tasks. For example, in a \textit{race condition}, tasks contend for exclusive write access to a single memory location and must \textit{synchronize} their reads to ensure correctness~\cite{McCool:2012}. While some dependencies can be refactored into a perfectly parallelizable form, others still require synchronization (e.g., locks and mutexes) or hardware \textit{atomic} primitives to prevent non-deterministic output. The key to enabling scalability in this scenario is to avoid high contention at any given memory location and prevent \textit{blocking} of tasks, whereby tasks remains idle (sometimes deadlocked) until they can access a lock resource. To enable lock-free progress of work among tasks, fine-grained atomic primitives are commonly used to efficiently check and increment values at memory locations~\cite{Herlihy:1991,Dice:2013}. For example, the \textit{compare-and-swap} (CAS) primitive atomically compares the value read at a location to an expected value. If the values are equal, then a new value is set at the location; otherwise, the value doesn't change.

Moreover, programs that have a high ratio of memory accesses to arithmetic computations can incur significant \textit{memory latency}, which is the number of clock or instruction cycles needed to complete a single memory access~\cite{Patterson:2008}. During this latency period, processors should perform a sufficient amount of parallel work to \textit{hide} the latency and avoid being idle. Given the \textit{bandwidth}, or instructions completed per cycle, of each processor, \textit{Little's Law} specifies the number of parallel instructions needed to hide latency as the bandwidth multiplied by latency~\cite{Little:2011}. While emerging many-core and massively-threaded architectures provide more available parallelism and higher bandwidth rates, the memory latency rate remains stagnant due to physical limitations~\cite{McCool:2012}. Thus, to exploit this greater throughput and \textit{instruction-level parallelism} (ILP), a program should ideally be decomposed into fine-grained units of computation that perform parallelizable work (\textit{fine-grained parallelism}).

Furthermore, the increase in available parallelism provided by emerging architectures also enables larger workloads and data to be processed in parallel~\cite{McCool:2012,Jeffers:2015}. Gustafson~\cite{Gustafson:1988} noted that as a problem size grows, the amount of parallel work increases much faster than the amount of serial work. Thus, a speedup can be achieved by decreasing the serial fraction of the total work. By explicitly parallelizing fine-grained computations that operate on this data, scalable \textit{data-parallelism} can be attained, whereby a single instruction is performed over multiple data elements (SIMD) in parallel (e.g., via a vector instruction), as opposed to over a single scalar data values (SISD). This differs from \textit{task-parallelism}, in which multiple tasks of a program conduct multiple instructions in parallel over the same data elements (MIMD)~\cite{Patterson:2008}. Task-parallelism only permits a constant speedup and induces \textit{coarse-grained parallelism}, whereby all tasks work in parallel but an individual task could still be executing serial work. By performing inner fine-grained parallelism within outer course-grained parallel tasks, a \textit{nested parallelism} is attained~\cite{Blikberg:2005}. Many recursive and segmented problems (e.g., quicksort and closest pair) can often be refactored into nested-parallel versions~\cite{blelloch1990vector}. Flynn~\cite{Flynn:1972} introduces SIMD, SISD, and MIMD as part of a taxonomy of computer instruction set architectures.
    
\subsection{General-Purpose Computing on GPU (GPGPU)}

%New architectures have unified memory (for hybrid general purpose computing), dynamic dispatch (for nested parallelism, inner recursion), improved memory bandwidth and latency hiding (of non-coalesced memory instructions), and cooperative groups for flexible synchronization and atomic aggregation (e.g. reduction or prefix sums) among varying sizes of threads groups, improved hardware for application-specific workloads, such as deep neural networks.
 
A \textit{graphical processing unit} (GPU) is a special-purpose architecture that is designed specifically for high-throughput, data-parallel computations that possess a high arithmetic intensity---the ratio of arithmetic operations to memory operations~\cite{Patterson:2008}. Traditionally used and hard-wired for accelerating computer graphics and image processing calculations, modern GPUs contain many times more execution cores and available instruction-level parallelism (ILP) than a CPU of comparable size~\cite{Nvidia-ProgramGuide}. This inherent ILP is provided by a group of processors, each of which performs SIMD-like instructions over thousands of independent, parallel threads. These \textit{stream processors} operate on sets of data, or \textit{streams}, that require similar computation and exhibit the following characteristics~\cite{Kapasi:2003}:
\begin{tightItemize}
\item \textit{High Arithmetic Intensity}: High number of arithmetic instructions per memory instruction. The stream processing should be largely compute-bound as opposed to memory bandwidth-bound.
\item \textit{High Data-Parallelism}: At each time step, a single instruction can be applied to a large number of streams, and each stream is not dependent on the results of other streams.
\item \textit{High Locality of Reference}: As many streams as possible in a set should align their memory accesses to the same segment of memory, minimizing the number of memory transactions to service the streams.   
\end{tightItemize}

\textit{General-purpose GPU} (GPGPU) computing leverages the massively-parallel hardware capabilities of the GPU for solving general-purpose problems that are traditionally computed on the CPU (i.e., non-graphics-related calculations). These problems should feature large data sets that can be processed in parallel and satisfy the characteristics of stream processing outlined above. Accordingly, algorithms for solving these problems should be redesigned and optimized for the data-parallel GPU architecture, which has significantly different hardware features and performance goals than a modern CPU architecture~\cite{Nvidia-BestPractices}.

Modern GPGPUs with dedicated memory are most-commonly packaged as discrete, programmable devices that can be added onto the motherboard of a compute system and programmed to configure and execute parallel functions~\cite{Patterson:2008}. The primary market leaders in the design of discrete GPGPUs are Nvidia and Advanced Micro Devices (AMD), with their GeForce and Radeon family of generational devices, respectively. Developed by Nvidia, the CUDA parallel programming library provides an interface to design algorithms for execution on an Nvidia GPU and configure hardware elements~\cite{Nvidia-ProgramGuide}. For the remainder of this survey, all references to a GPU will be with respect to a modern Nvidia CUDA-enabled GPU, as it is used prevalently in most of the GPU hashing studies.  

The following subsections review important features of the GPU architecture and discuss criteria for optimal GPU performance.
%, and outline how these criteria differ, at a high level, from a CPU architecture.

\subsubsection{SIMT Architecture}
A GPU is designed specifically for \textit{Single-Instruction, Multiple Threads} (SIMT) execution, which is a combination of SIMD and simultaneous multi-threading (SMT) execution that was introduced by Nvidia in 2006 as part of the Tesla micro-architecture~\cite{Nvidia-PTX}. On the host CPU, a program, or \textit{kernel} function, is written in CUDA C and invoked for execution on the GPU. The kernel is executed $N$ times in parallel by $N$ different CUDA threads, which are dispatched as equally-sized \textit{thread blocks}. The total number of threads is equal to the number of thread blocks times the number of threads per block, both of which are user-defined in the kernel. Thread blocks are required to be independent and can be scheduled in any order to be executed in parallel on one of several independent \textit{streaming multi-processors} (SMs). The number of blocks is typically based on the number of data elements being processed by the kernel or the number of available SMs~\cite{Nvidia-ProgramGuide}. Since each SM has limited memory resources available for resident thread blocks, there is a limit to the number of threads per block---typically $1024$ threads. Given these memory constraints, all SMs may be occupied at once and some thread blocks will be left inactive. As thread blocks terminate, a dedicated GPU scheduling unit launches new thread blocks onto the vacant SMs.

Each SM chip contains hundreds of ALU (arithmetic logic unit) and SFU (special function unit) compute cores and an interconnection network that provides $k$-way access to any of the $k$ partitions of off-chip, high-bandwidth global DRAM memory. Memory requests first query a global L2 cache and then only proceed to global memory upon a cache miss. Additionally, a read-only texture memory space is provided to cache global memory data and enable fast loads. On-chip thread management and scheduling units pack each thread block on the SM into one or more smaller logical processing groups known as \textit{warps}---typically $32$ threads per warp; these warps compose a \textit{cooperative thread array} (CTA). The thread manager ensures that each CTA is allocated sufficient shared memory space and per-thread registers (user-specified in kernel program). This on-chip shared memory is designed to be low-latency near the compute cores and can be programmed to serve as L1 cache or different ratios thereof (newer generations now include these as separate memory spaces)~\cite{Nvidia-Maxwell}.

Finally, each time an instruction is issued, the SM instruction scheduler selects a warp that is ready to execute the next SIMT scalar (register-based) instruction, which is executed independently and in parallel by each active thread in the warp. In particular, the scheduler applies an active mask to the warp to ensure that only active threads issue the instruction; individual threads in a warp may be inactive due to independent branching in the program. A synchronization barrier detects when all threads (and warps) of a CTA have exited and then frees the warp resources and informs the scheduler that these warps are now ready to process new instructions, much like context switching on the CPU. Unlike a CPU, the SM does not perform any branch prediction or speculative execution (e.g., prefetching memory) among warp threads~\cite{Patterson:2008}.

SIMT execution is similar to SIMD, but differs in that SIMT applies one instruction to multiple independent warp threads in parallel, instead of to multiple data lanes. In SIMT, scalar instructions control individual threads, whereas in SIMD, vector instructions control the entire set of data lanes. This detachment from the vector-based processing enables threads of a warp to conduct a form of SMT execution, where each thread behaves more like a heavier-weight CPU thread~\cite{Patterson:2008}. Each thread has its own set of registers, addressable memory requests, and control flow. Warp threads may take divergent paths to complete an instruction (e.g., via conditional statements) and contribute to starvation as faster-completing threads wait for the slower threads to finish.

The two-level GPU hierarchy of warps within SMs offers massive nested parallelism over data~\cite{Patterson:2008}. At the outer, SM level of granularity, coarse-grained parallelism is attained by distributing thread blocks onto independent, parallel SMs for execution. Then at the inner, warp level of granularity, fine-grained data and thread parallelism is achieved via the SIMT execution of an instruction among parallel warp threads, each of which operates on an individual data element. The massive data-parallelism and available compute cores are provided specifically for high-throughput, arithmetically-intense tasks with large amounts of data to be independently processed. If a high-latency memory load is made, then it is expected that the remaining warps and processors will simultaneously perform sufficient work to hide this latency; otherwise, hardware resources remain unused and yield a lower aggregate throughput~\cite{Volkov:2016}. The GPU design trades-off lower memory latency and larger cache sizes (such as on a CPU) for increased instruction throughput via the massive parallel multi-threading~\cite{Patterson:2008}.

This architecture description is based on the Nvidia Maxwell micro-architecture, which was released in 2015~\cite{Nvidia-Maxwell}. While certain quantities of components (e.g., SMs, compute cores, memory sizes, and thread block sizes) change with each new generational release of the Nvidia GPU, the general architectural design and execution model remain constant~\cite{Nvidia-ProgramGuide}. The CUDA C Programming Guide~\cite{Nvidia-ProgramGuide} and Nvidia PTX ISA documentation~\cite{Nvidia-PTX} contain further details on the GPU architecture, execution and memory models, and CUDA programming.

\subsubsection{Optimal Performance Criteria}

The following performance strategies are critical for maximizing utilization, memory throughput, and instruction throughput on the GPU~\cite{Nvidia-BestPractices}.

\textit{Sufficient parallelism}: Sufficient instruction-level and thread-level parallelism should be attained to fully hide arithmetic and memory latencies.  According to Little's Law, the number of parallel instructions needed to hide a latency (number of cycles needed to perform an instruction) is roughly the latency times the throughput (number of instructions performed per cycle)~\cite{Little:2011}. During this latency period, threads that are dependent on the output data of other currently-executing threads in a warp (or thread block) are stalled. Thus, this latency can be hidden either by having these threads simultaneously perform additional, non-dependent SIMT instructions in parallel (instruction-level parallelism), or by increasing the number of concurrently running warps and warp threads (thread-level parallelism)~\cite{Volkov:2016}.

Since each SM has limited memory resources for threads, the number of concurrent warps possible on an SM is a function of several configurable components: allocated shared memory, number of registers per thread, and number of threads per thread block~\cite{Nvidia-ProgramGuide}. Based on these parameters, the number of parallel thread blocks and warps on an SM can be calculated and used to compute the \textit{occupancy}, or ratio of the number of active warps to the maximum number of warps. In terms of Little's Law, sufficient parallel work can be exploited with either a high occupancy or low occupancy, depending on the amount of work per thread. Based on the specific demands for SM resources, such as shared memory or register usage, by the kernel program, the number of available warps will vary accordingly. Higher occupancy, usually past $50$ percent, does not always translate into improved performance~\cite{Nvidia-BestPractices}. For example, a lower occupancy kernel will have more registers available per thread than a higher occupancy kernel, allowing low-latency access to local variables and minimizing register spilling into high-latency local memory. 

%On a modern GPU, a typical arithmetic instruction has a low latency and large throughput for a compute-bound, arithmetically-intense kernel programs. Thus, a high instruction-level parallelism can be achieved with a lower occupancy, or fewer active warps. However, a global memory-bound instruction typically has a much higher latency and thus
%Latency is the number of processor clock cycles required to perform a given instruction.
%On a current CUDA-enabled Nvidia GPU, a typical arithmetic instruction has a latency of approximately 24 cycles, while a memory-bound instruction has a latency of more than 400.
%Whereas latency measures the time (cycles) needed per instruction, throughput measures the rate of instructions completed per cycle.
%A common fallacy is that a thread can access shared memory just as fast as a hardware register.
%One of several factors that determine occupancy is register availability. Register storage enables threads to keep local variables nearby for low-latency access. However, the set of registers is a limited commodity that all threads resident on a multiprocessor must share. Registers are allocated to an entire block all at once. So, if each thread block uses many registers, the number of thread blocks that can be resident on a multiprocessor is reduced, thereby lowering the occupancy of the multiprocessor.

\textit{Memory coalescing}: When a warp executes an instruction that accesses global memory, it \textit{coalesces} the memory accesses of the threads within the warp into one or more memory transactions, or cache lines, depending on the size of the word accessed by each thread and the spatial coherency of the requested memory addresses. To minimize transactions and  maximize memory throughput, threads within a warp should coherently access memory addresses that fit within the same cache line or transaction. Otherwise, memory divergence occurs and multiple lines of memory are fetched, each containing many unused words. In the worst case alignment, each of the 32 warp threads accesses successive memory addresses that are multiples of the cache line size, prompting 32 successive load transactions~\cite{Nvidia-BestPractices}. 

The shared memory available to each thread block can help coalesce or eliminate redundant accesses to global memory~\cite{Patterson:2008}. The threads of the block (and associated warp) can share their data and coordinate memory accesses to save significant global memory bandwidth. However, it also can act as a constraint on SM occupancy---particularly limiting the number of available registers per thread and warps---and is prone to \textit{bank conflicts}, which occur when two or more threads in a warp access an address in the same bank, or partition, of shared memory~\cite{Nvidia-ProgramGuide}. Since an SM only contains one hardware bus to each bank, multiple requests to a bank must be serialized. Thus, optimal use of shared memory necessitates that warp threads arrange their accesses to different banks~\cite{Nvidia-ProgramGuide}. Finally, the read-only texture memory of an SM can be used by a warp to perform fast, non-coalesced lookups of cached global memory, usually in smaller transaction widths.  

\textit{Control flow}: Control flow instructions (e.g., if, switch, do, for, while) can significantly affect instruction throughput by causing threads of the same warp to diverge and follow different execution paths, or branches. Optimal control flow is realized when all the threads within a warp follow the same execution path~\cite{Nvidia-BestPractices}. This scenario enables SIMD-like processing, whereby all threads complete an instruction simultaneously in lock-step. During \textit{branch divergence} in a warp, the different executions paths, or branches, must be serialized, increasing the total number of instructions executed for the warp. Additionally, the use of atomics and synchronization primitives can also require additional serialized instructions and thread starvation within a warp, particularly during high contention for updating a particular memory location~\cite{Stuart:2011}.

\subsection{Data Parallel Primitives}

The redesign of serial algorithms for scalable data-parallelism offers platform portability, as increases in processing units and data are accompanied by unrestricted increases in speedup. \textit{Data-parallel primitives} (DPPs) provide a way to explicitly design and program an algorithm for this scalable, platform-portable data-parallelism. DPPs are highly-optimized \textit{building blocks} that are combined together to compose a larger algorithm. The traditional design of this algorithm is thus refactored in terms of DPPs. By providing highly-optimized implementations of each DPP for each platform architecture, an algorithm composed of DPPs can be executed efficiently across multiple platforms. This use of DPPs eliminates the combinatorial (cross-product) programming issue of having to implement a different version of the algorithm for each different architecture.

The early work on DPPs was set forth by Blelloch~\cite{blelloch1990vector}, who proposed a \textit{scan vector model} for parallel computing. In this model, a vector-RAM (V-RAM) machine architecture is composed of a vector memory and a parallel vector processor. The processor executes vector instructions, or \textit{primitives}, that operate on one or more arbitrarily-long vectors of atomic data elements, which are stored in the vector memory. This is equivalent to having as many independent, parallel processors as there are data elements to be processed. Each primitive is classified as either \textit{scan} or \textit{segmented} (per-segment parallel instruction), and must possess a parallel, or \textit{step}, time complexity of $O(\log n)$ and a serial, or \textit{element}, time complexity of $O(n)$, in terms of $n$ data elements; the element complexity is the time needed to simulate the primitive on a serial random access machine (RAM). Several canonical primitives are then introduced and used as building blocks to refactor a variety of data structures and algorithms into data-parallel forms.

The following are examples of DPPs that are commonly-used as building blocks to construct data-parallel algorithms: 
\begin{tightItemize}
\item \textit{Map}: Applies an operation on all elements of the input array, storing the result in an output array of the same size, at the same index;
\item \textit{Reduce}: Applies an aggregate binary operation (e.g., summation or maximum) on all elements of an input array, yielding a single output value. 
\textit{ReduceByKey} is a variation that performs segmented \textit{Reduce} on the input array based on unique key, yielding an output value for each key;
\item \textit{Gather}: Given an input array of values, reads values into an output array according to an array of indices;
\item \textit{Scan}: Calculates partial aggregates, or a prefix sum, for all values in an input array and stores them in an output array of the same size;
\item \textit{Scatter}: Writes each value of an input data array into an index in an output array, as specified in the array of indices;
\item \textit{Compact}: Applies a unary predicate (e.g., if an input element is greater than zero) on all values in an input array, filtering out all the values which do not satisfy the predicate. Only the remaining elements are copied into an output array of an equal or smaller size;
\item \textit{SortByKey}: conducts an in-place segmented \textit{Sort} on the input array, with segments based on a key or unique data value in the input array;
\item \textit{Unique}: Ignores duplicate values which are adjacent to each other, copying only unique values from the input array to the output array of the same or lesser size; and
\item \textit{Zip}: Binds two arrays of the same size into an output array of pairs, with the first and second components of a pair equal to array values at a given index.
\end{tightItemize}
Several other DPPs exist, each meeting the required step and element complexities specified by Blelloch~\cite{blelloch1990vector}. Cross-platform implementations of a wide variety of DPPs form the basis of several notable open-source libraries.

The Many-Core Visualization Toolkit (VTK-m)~\cite{Moreland:CGA2016} is a platform-portable library that provides a growing set of DPPs and DPP-based algorithms~\cite{vtk-m}. With a single code base, back-end code generation and runtime support are provided for use on GPUs and CPUs. Currently, each GPU-based DPP is a modified variant from the Nvidia CUDA Thrust library of parallel algorithms and data structures~\cite{Thrust:web:2017}, and each CPU-based DPP is adopted from the Intel Thread Building Blocks (TBB) library for scalable data parallel programming~\cite{TBB:web:2017}. VTK-m provides the flexibility to develop custom \textit{device adapter algorithms}, or DPPs, for a new device type. This device can take the form of an emerging architecture or a new parallel programming language (e.g., Thrust and TBB) for which DPPs must be re-optimized. Thus, a DPP can be invoked in the high-level VTK-m user code and executed on any of the devices at runtime. The choice of device is either specified at compile-time by the user, or automatically selected by VTK-m. VTK-m, Thrust, and TBB all employ a generic programming model that provides C++ Standard Template Library (STL)-like interfaces to DPPs and algorithms~\cite{Plauger:2000}. Templated arrays form the primitive data structures over which elements are parallelized and operated on by DPPs. Many of these array types provide additional functionality on top of underlying vector iterators that are inspired by those in the Boost Iterator Library~\cite{BoostIterator:web:2003}. 

%VTK-m is effectively the unification of three predecessor visualization libraries---DAX~\cite{DAX}, EAVL~\cite{EAVL}, and PISTON~\cite{lo2012piston}---each of which is built upon DPPs, with an aim to achieve portable performance across multiple many-core architectures. Moreland et al.~\cite{Moreland:2013} categorize scientific visualization algorithms based on their data-parallel design patterns and then propose DPP-based re-factorings of these algorithms that exploit the massively-parallel threading available on emerging processors and accelerators.

The CUDA Data Parallel Primitives Library (CUDPP)~\cite{cuddp} is a library of fundamental DPPs and algorithms written in Nvidia CUDA C~\cite{Nvidia-ProgramGuide} and designed for high-performance execution on CUDA-compatible GPUs. Each DPP and algorithm incorporated into the library is considered best-in-class and typically published in peer-reviewed literature (e.g., radix sort~\cite{Merrill:2010,Ashkiani:2016}, mergesort~\cite{Satish:2009,Davidson:2012}, and cuckoo hashing~\cite{Alcantara:2009,Alcantara:2011}). Thus, its data-parallel implementations are constantly updated to reflect the state-of-the-art.

\subsection{GPU Searching}
\label{gpu-data-structures}
The following section reviews canonical approaches for organizing, storing, and searching data on the GPU. 

Let $U = \{i\}_{0 \leq i < u}$ be the universe for some arbitrary positive integer $u$. Then let $S \subset U$ be an unordered set of $n = |S|$ elements, or \textit{keys}, belonging to $U$. The \textit{search problem} seeks an answer to the \textit{query}: ``Is key $k$ a member of $S$?'' If $k \in S$, then we return its corresponding \textit{value}, which is either $k$ itself or a different value. A \textit{data structure} is built or constructed over $S$ to efficiently facilitate the searching operation. The data structure is implementation-specific and can be as simple as a sorted (ordered) variant of the original set, a hash table, or a tree-based partitioning of the elements.
    
A generalization of the search task is the \textit{dictionary problem}, which seeks to both modify and query key-value pairs $(k,v)$ in $S$. A canonical dictionary data structure supports $insert(k,v)$, $delete(k,v)$, $query(k)$, $range(k_1,k_2)$ (returns $\{k|k_1 \leq k \leq k_2\}$), and $count(k_1,k_2)$ (returns $|range(k_1,k_2)|$). To support these operations, the dictionary must be dynamic and accommodate incremental or batch updates after construction; this contrasts to a static data structure, which either does not support updates after a one-time build or must be rebuilt after each update. In multi-threaded environments, these structures must also provide concurrency and ensure correctness among mixed, parallel operations that may access the same elements simultaneously. 

An extensive body of work has embarked on the redesign of data structures for construction and general computation on the GPU~\cite{Owens:2007:Survey}. Within the context of searching, these \textit{acceleration} structures include sorted arrays~\cite{Alcantara:2009,Schlegel:2009,Alcantara:2011,Kaldewey:2012,Lessley:EGPGV2016,Lessley:LDAV17-1,Ashkiani:LFAO17} and linked lists~\cite{Yang:2010}, hash tables (see section~\ref{hashing-techniques}), spatial-partitioning trees (e.g., $k$-d trees~\cite{Zhou:2008,Karras:2012,Widanagamaachchi:2014}, octrees~\cite{Zhou:2011,Karras:2012}, bounding volume hierarchies (BVH)~\cite{Lauterbach:2009,Karras:2012}, R-trees~\cite{Luo:2012}, and binary indexing trees~\cite{Kim:2010,Schneider:2017}), spatial-partitioning grids (e.g., uniform~\cite{Lagae:2008,Kalojanov:2009,Gao:2015} and two-level~\cite{Kalojanov:2011}), skiplists~\cite{Moscovici:2017}, and queues (e.g., binary heap priority~\cite{He:2012} and FIFO~\cite{Cederman:2012,Scogland:2015}). Due to significant architectural differences between the CPU and GPU, search structures cannot simply be ``ported'' from the CPU to the GPU and maintain optimal performance. On the CPU, these structures can be designed to fit within larger cache, perform recursion, and employ heavier-weight synchronization or hardware atomics. However, during queries, the occurrence of varying paths of pointers (pointer chasing) and dependencies between different phases or levels of the structure both limit the parallel throughput on the GPU. Moreover, these structures ideally should be constructed directly on the GPU, as transfers from the CPU over the PCIe bus induce costly latencies. 

For searching an unordered array of elements on the GPU, two canonical data structures exist: the sorted array and the hash table. Both of these data structures are known to be relatively fast to construct on the GPU and are amenable to data-parallel design patterns~\cite{Ashkiani:LFAO17}.
    
\subsubsection{Searching Via Sorting}
Given a set of $n$ unordered elements, a canonical searching approach is to first sort the elements in ascending order and then conduct a binary or $k$-nary search for the query element. This search requires a logarithmic number of comparisons in the worst-case, but is not as amenable to caching as consecutive comparisons are not spatially close in memory for large $n$. Moreover, on the GPU, an ordered query pattern by threads in a warp can enable memory coalescing during comparisons. 

The current version of the CUDA Thrust library~\cite{Thrust:web:2017} provides fast and high-throughput data-parallel implementations of mergesort~\cite{Satish:2009} and radix sort~\cite{Merrill:2010} for arrays of custom (e.g., comparator function) or numerical (i.e., integers and floats) data types, respectively. Similarly, the latest version of the CUDPP library~\cite{cuddp} includes best-in-class data-parallel algorithms for mergesort~\cite{Satish:2009,Davidson:2012} and radix sort~\cite{Merrill:2010,Ashkiani:2016}, each of which are adapted from published work. Singh et al.~\cite{Singh:2017} survey and compare the large body of recent GPU-based sorting techniques.

A few studies have investigated various factors that affect the performance of data-parallel sort methods within the context of searching~\cite{Alcantara:2009,Alcantara:2011,Lessley:LDAV17-1}. Kaldewey and Blas introduce a GPU-based $p$-ary search that first uses $p$ parallel threads to locate a query key within one of $p$ larger segments of a sorted array, and then iteratively repeats the procedure over $p$ smaller segments within the larger segment. This search achieves high memory throughput and is amenable to memory coalescing among the threads~\cite{Kaldewey:2012}. Moreover, the algorithm was also ported to the CPU to leverage the SIMD vector instructions in a fashion similar to the $k$-ary search introduced by Schlegel et al.~\cite{Schlegel:2009}. However, the fixed vector width restricts the degree of parallelism and value of $p$, which is significantly higher on the GPU.

Inserting or deleting elements into a sorted array is generally not supported and requires inefficient approaches such as appending/removing new elements and re-sorting the larger/smaller array, or first sorting the batch of new insertions and then merging them into the existing sorted array.  Ashkiani et al.~\cite{Ashkiani:LFAO17} present these approaches and the resulting performance for a dynamic sorting-based dictionary data structure, along with setting forth the current challenges of designing dynamic data structures on the GPU.
    
\subsubsection{Searching Via Hashing}
\label{hashing}

Instead of maintaining elements in sorted order and performing a logarithmic number of lookups per query, \textit{hash tables} compactly reorganize the elements such that only a constant number of direct, random-access lookups are needed on average~\cite{Cormen:2001}. More formally, given a universe $U$ of possible keys and an unordered set $S \subseteq U$ of $n$ keys (not necessarily distinct), a \textit{hash function}, $h: U \mapsto H$, maps the keys from $S$ to the range $H = \{j\}_{0 \leq j < m}$ for some arbitrary positive integer $m \geq n$. Defining a memory space over this range of size $m$ specifies a hash table, into which keys are inserted and queried. Thus, the hash table is addressable by the hash function. During an insertion or query operation for a key $q$, the hash function computes an address $h(q) = r$ into $H$. If the location $H[r]$ is empty, then $q$ is either inserted into $H[r]$ (for an insertion) or does not exist in $H$ (for a query). If $H[r]$ contains the key $q$ (for a query), then either $q$ or an associated \textit{value} of $q$ is returned\footnotemark, indicating success. Otherwise, if multiple distinct keys $q' \neq q$ are hashed to the same address $h(q') = r$, then a situation known as a hash \textit{collision} occurs. These collisions are typically resolved via \textit{separate chaining} (i.e., employing linked lists to store multiple keys at a single address) or \textit{open-addressing} (e.g., when an address is occupied, then store the key at the next empty address). 

\footnotetext{In practice, the values should be easily stored and accessible within an auxiliary array or via a custom arrangement within the hash table.}

The occurrence of collisions deteriorates the query performance, as each of the collided keys must be iteratively inspected and compared against the query key. According to the birthday paradox, with a discrete uniform distribution hash function that outputs a value between 1 and 365 for any key, the probability that two random keys hash to the same address in a hash table of size 23 is 50 percent~\cite{Suzuki:2006}. More generally, for $n$ hash values and a table size of $m$, the probability $p(n,m)$ of a collision is
\begin{align*}
p(n,m) &= \begin{cases} 1-\displaystyle\prod_{k=1}^{n-1}\left(1-\frac{k}{m}\right) & n\le m \\ 1 & n > m \end{cases} \\[8px]
& \approx 1 - \left( \frac{m-1}{m} \right)^\frac{n(n-1)}{2}.
\end{align*}
Thus, for a large number of keys ($n$) and small hash table ($m$), hash collisions are inevitable.

In order to minimize collisions, an initial approach is to use a good quality hash function that is both efficient to compute and distributes keys as evenly as possible throughout the hash table~\cite{Cormen:2001}. One such family of functions are randomly-generated, parameterized functions of the form $h(k) = (a\cdot k + b)\bmod p\bmod |H|$, where $p$ is a large prime number and $a$ and $b$ are randomly-generated constants that bias $h$ from outputting duplicate values~\cite{Alcantara:2011}. However, $h$ is a function of the table size, $|H|$. If $|H|$ is too small, then not even the best of hash functions can avoid an increase in collisions. Given the table size, the \textit{load factor} $\alpha$ of the table is defined as $\alpha = n/|H|$, or the percentage of occupied addresses in the hash table, which $|H|$ is typically larger than $n$. If new keys are inserted into the table and $\alpha$ reaches a maximum threshold, then typically the table is allocated to a larger size and all the keys are \textit{rehashed} into the table.  

To avoid collision resolution altogether, a \textit{perfect} hash function can be constructed to hash keys into a hash table without collisions. Each key is mapped to a distinct address in the table. However, composing such a perfect hashing scheme is known to be difficult in general~\cite{Lefebvre:2006}. The probability of attaining a perfect hash for $n$ keys in a large table of size $m$ ($m \gg n$) is defined as
\begin{align*}
p(n,m) &= \left(1\right)\cdot \left(1 - \frac{1}{m}\right)\cdot \left(1 - \frac{2}{m}\right)\cdots \left(1 - \frac{n-1}{m}\right) \\
       &\approx e^{\frac{-n^2}{2m}},
\end{align*}
which is very small for a large $n$ or small $m$. Nonetheless, a significant body of research has investigated this approach and is reviewed in this survey, within the context of parallel hashing. 

A hash table is \textit{static} if it does not support modification after being constructed; that is, the table is only constructed to handle query operations. Thus, a static hash table also does not support mixed operations and the initial batch of insertions used to construct the table (bulk build) must be completed before the batch of query operations. A hash table that can be updated, or mutated, via insertion and deletion operations post-construction is considered \textit{dynamic}. Denoting the query, insert, and delete operations as $q$, $i$, and $d$, respectively, the \textit{operation distribution} $\Gamma = (q, i, d), q + i + d = 1$ specifies the percentage of each operation that are conducted concurrently in a hashing workload~\cite{Ashkiani:2017}. For example, $\Gamma = (0.7, 0.15, 0.15)$ represents a query-heavy workload that performs $70\%$ queries and $30\%$ updates. Additionally, the percentage $q$ can be split into queried keys that exist in the hash table and those that do not. Often, queries for non-existent keys can present worst-case scenarios for many hash techniques, as a maximum number of searches are conducted until failure~\cite{Ashkiani:2017}.   

As general data structures, hash tables do not place any special emphasis on the key access patterns over time~\cite{Choudhury:2016}. However, the patterns that appear in various real-world applications do possess observable structure. For example, geometric tasks may query spatially-close keys in a sequential or coherent pattern, and database tasks may query certain subsets of keys more frequently than others, whereby the hash table serves as a working set or most-recently-used (MRU) table for cache-like accesses~\cite{Alcantara:2011,Polychroniou:2013,Choudhury:2016}. Moreover, dynamic hash tables do not place special emphasis on the mixture $\Gamma$, or pattern, of query and update operations. However, execution time performance may be better or worse for some hashing techniques, depending on the specific $\Gamma$, such as query-heavy for key-value stores~\cite{Zhang:2015} or update-heavy for real-time, interactive spatial hash tables~\cite{Lefebvre:2006,Alcantara:2009,Garcia:2011,Niener:2013}.

Finally, hash tables offer compact storage for sparse spatial data that contains repeated elements or empty elements that don't need to be computed. For example, instead of storing an entire, mostly-empty voxelized 3D grid, the non-empty voxels can be hashed into a dense hash table~\cite{Lefebvre:2006}. Then, every voxel can be queried to determine whether it should be rendered or not, returning a negative result for the empty voxels. Furthermore, a hash table does not have to be one-dimensional. Instead, the data structure can consist of multiple hash tables or \textit{bucketed} partitions that are each addressed by a different hash function.

While collision resolution is straightforward to implement in a serial CPU setting, it does not easily translate to a parallel setting, particularly on massively-threaded, data-parallel GPU architectures. GPU-based hashing\footnotemark presents several notable challenges:
\begin{tightItemize}
\item Hashing is a memory-bound problem that is not as amenable to the compute-bound and limited-caching design of the GPU, which hides memory latencies via a large arithmetic throughput.
\item The random-access nature of hashing can lead to disparate writes and reads by parallel-cooperating threads on the GPU, which performs best when memory accesses are coalesced or spatially coherent.
\item The limited memory available on a GPU puts restrictions on the maximum hash table size and number of tables that can reside on device.
\item Collision resolution schemes handle varying numbers of keys that are hashed and chained to the same address (separate chaining), or varying numbers of attempts to place a new, collided key into an empty table location (open-addressing). This variance causes some insert and query operations to require more work than others. On a GPU, threads work in groups to execute the same operation on keys in a data-parallel fashion. Thus, a performance bottleneck arises when faster, non-colliding threads wait for slower, colliding threads to finish. Moreover, some threads may insert colliding keys that are unable to find an empty table location, leading to failure during construction of the table.
\end{tightItemize}

Searching via the construction and usage of a hash table on the GPU has recently received a breadth of new research, with a variety of different parallel designs and applications, ranging from collision detection to surface rendering to nearest neighbor approximation. The following section covers these GPU-based parallel hashing approaches.

\footnotetext{In this study, the term \textit{hashing} refers to the entire process from constructing the hash table to the handling of collisions while querying or updating; this is not to be confused with hash function design and computation, or its application to cryptographic protocols and message passing.}

%Even with a worst-case constant number of probes, the number of probes per query may still vary, leaving the branch divergence issue unresolved.
%A possible direction for addressing this is to maintain spatial coherence among spatially-close elements in the hash table.
%This is accomplished by hashing neighboring elements to neighboring locations in the hash table.
%Thus, within a warp, a single memory load instruction can service the coherent queries of each of the threads, leading to memory coalescing and a similar number of probes for each thread, which alleviates the divergence bottleneck of the SIMD parallelism.
\section{Hashing Techniques}
\label{hashing-techniques}

We consider four different categories of hashing techniques: open-addressing probing, perfect hashing, spatial hashing, and separate chaining. Each category is discussed in a separate subsection and distinguished by its method of handling hash collisions or placement of elements within the hash table.

\subsection{Open-addressing Probing}

In \textit{open-addressing}, a key is inserted into the hash table by \textit{probing}, or searching, through alternate table locations---the \textit{probe sequence}---until a location is found to place the element~\cite{Cormen:2001}. The determination of where to place the element varies by probing scheme: some schemes probe for the first unused location (\textit{empty slot}), whereas others \textit{evict} the currently-residing key at the probe location (i.e., a collision) and swap in the new key. Each probe location is specified by a hash function unique to the probing scheme. Thus, some probe sequences may be more compact or greater in length than others, depending on the probing method. For a query operation, the locations of the probe sequence are computed and followed to search for the queried key in the table.  

Each probing method trades-off different measures of performance with respect to GPU-based hashing. A critical influence on performance is the load factor, which is the percentage of occupied locations in the hash table (subsection~\ref{hashing}). As the load factor increases towards 100 percent, the number of probes needed to insert or query a key increases greatly. Once the table becomes full, probing sequences may continue indefinitely, unless bounded, and lead to insertion failure and possibly a hashing \textit{restart}, whereby the hash table is reconstructed with different hash functions and parameters. Moreover, for threads within a warp on the GPU, variability in the number of probes per thread can induce branch divergence and inefficient SIMD parallelism, as all the threads will need to wait for the worst-case number of probes to execute the next instruction. 

The following subsections review research on open-addressing probing for GPU-based hashing, distinguishing each study by its general probing scheme: linear probing, cuckoo hashing, double hashing, multi-level or bucketized probing, and robin hood hashing.

\subsubsection{Linear Probing-based Hashing}
\label{linear-probing}

\textit{Linear probing} is the most basic method of open-addressing. In this method, a key $k$ first hashes to location $h(k)$ in the hash table. Then, if the location is already occupied, $k$ linearly searches locations $h(k) + 1, h(k) + 2,\dots$etc. until an empty slot (insertion) or $k$ itself (query) is found. If $h(k)$ is empty, then $k$ is inserted immediately, without probing; otherwise, a worst-case $O(n)$ probes will need to be made to locate $k$ or an empty slot, where $n$ is the size of the hash table. While simple in design, linear probing suffers from \textit{primary clustering}, whereby a cluster, or contiguous block, of locations following $h(k)$ are occupied by keys, reducing nearby empty slots. This occurs because colliding keys at $h(k)$ each successively probe for the next available empty slot after $h(k)$ and insert themselves into it. An improved variant of linear probing is \textit{quadratic probing}, which replaces the linear probe sequence starting at $h(k)$ with successive values of an arbitrary quadratic polynomial: $h(k) + 1^2, h(k) + 2^2,\dots$etc. This avoids primary clustering, but also introduces a \textit{secondary clustering} effect as a result. For a more than half-full table, both of these probing methods can incur a long probe sequence to find an empty slot, possibly resulting in failure during an insert.

Bordawekar~\cite{Bordawekar:2014} develops an open-addressing approach based on multi-level bounded linear probing, where the hash table has multiple levels to reduce the number of lookups during linear probing. In the first level hash table, each key hashes to a location $h_{1}(k)$ and then looks for an empty location, via linear probing, within a bounded probe region $P_{1} = [h_{1}(k), h_{1}(k)+(j-1)]$, where $j$ is the size of the region. If an empty location is not found, then the key must be inserted into the second-level hash table, which is accomplished by hashing to location $h_2(k)$ and linear probing within another, yet larger, probe region $P_{2}$. This procedure continues for each level, until an empty location is found. In this work, only 2-level and 3-level hash tables are considered; thus, a thread must perform bounded probing on a key for at most three rounds, before declaring failure. To query a key, a thread completes the same hashing and probing procedure. In a data-parallel fashion, each thread within a warp is assigned a key from the bounded probe region and compares this key with the query key, using warp-level voting to communicate success or failure. This continues across warps, for each hash table level.

The initial design goal of this multi-level approach was to bound and reduce the average number of probes per insertion and query, while enabling memory coalescing and cache line coherency among threads (or lanes) within a warp. By using a small, constant number of hash tables and functions, the load factor could be increased beyond the $70$ percent of Alcantara et al.'s cuckoo hashing (subsection~\ref{cuckoo-hashing}), without sacrificing performance. However, experimental results reveal that this approach, with both two and three levels (and hash functions), does not perform as fast as cuckoo hashing for the largest batches of key-value pairs (hundreds of millions); for smaller batches, the multi-level approaches are the best performers. This finding is particularly noticeable for querying the keys, suggesting that improved probing and memory coalescing are likely not achieved. Additional details are needed to ascertain whether the ordering of the keys---spatial or random---affect this multi-level approach, or specific reasons why the expected warp-level memory coalescing is not being realized.

Karnagel et al.~\cite{Karnagel:2015} develop a linear probing hashing scheme to perform database group-by and aggregation queries (i.e., a reduce-by-key operation) on the GPU.
In this work, a database of records (e.g., customer data) is stored in SSDs and then queried in SQL format from the CPU. Selected columns of the query (e.g., zip code and order total) are transferred and pinned into host memory, from which the GPU fetches the column values in a coalesced, data-parallel fashion via Universal Virtual Addressing (UVA). Then, a hashing procedure begins to compute an aggregate (e.g. average discount) for each unique item, or \textit{group}, in the group-by column (e.g., customer zip code). Each item in this column is initially inserted into a global memory hash table as a key-value pair, where the key is the item ID and the value is a tuple (count, sum). For each item, the ID key is hashed to a location and then probes linearly until its matching key is found. If an empty slot is encountered, then a new value tuple is inserted for the key; otherwise, the count and sum of the tuple are atomically added to the current value tuple for the key. After all threads have completed, the count and sum of each key in the hash table are used to calculate aggregate values. These aggregate values form the output of the query. All GPU hash table operations and arithmetic computations are performed in data-parallel fashion by blocks of threads.

The primary contribution of this work is a thorough experimental evaluation of several factors affecting hashing performance on the GPU, including guidance on how to decide the hash table size, load factor, and CUDA grid configuration of number of thread blocks and threads per block. Notable experimental findings are the following: 
\begin{itemize}
\item As the number of groups increases, either the hash table must grow proportionately in size, or the load factor needs to increase. The ideal table size is one that fits within shared L2 cache, with a load factor below 50 percent. If a higher load factor is used, then thread contention and long probe sequences emerge. This contention is due to multiple threads attempting to access the value tuple for the same key (group) and having to synchronize via atomic compare-and-swap updates. 
\item Hash tables that cannot fit within L2 cache reside in global memory and each thread must make global memory accesses, unless the data is cached. For the data to reside in cache, a cache line load of a small, fixed size (e.g., 128 bytes for L1 and 32 bytes for L2) must be performed. Since linear probing creates a variable number of probes per thread and threads access random, non-coalesced regions of memory, a thread will likely need multiple cache line loads to complete an operation. With thousands of concurrent threads, each invoking cache line loads into a limited size cache, the chances of cache pollution and eviction are high, diminishing the benefits of caching. Thus, to achieve undisturbed cache access to data, fewer threads should be used. The number of threads, however, should also be at least enough to hide the memory latency of the PCIe data tranfers from host to device.  
\end{itemize}
Additionally, the simple hash table and probing scheme are only used in order to minimize the number of factors affecting performance and because the approach is mainly PCIe-bandwidth bound, which affords more probes and non-coalesced memory accesses to hide the latency. The authors acknowledge the bounded linear probing approach of Bordawekar~\cite{Bordawekar:2014}, but cite the latter reason for using a simpler hashing scheme.
%Also, the parameter values and memory sizes recommended in this study specifically pertain to the Nvidia GTX Titan GPU (Kepler generation) architecture. As cache, shared memory, and thread block capabilities change in emerging architectures, the trends observed should still hold and likely enable even more query throughput (number of groups to aggregate).

\subsubsection{Cuckoo-based Hashing}
\label{cuckoo-hashing}

In \textit{cuckoo hashing}, each key is assigned two locations in the hash table, as specified by primary and secondary hash functions~\cite{Pagh:2004}. When inserting a new key, its first location is probed with the primary function and the contents of the location are inspected. If the slot is empty, then the key is inserted and the probe sequence ends. Otherwise, a collided key already occupies the slot and the cuckoo eviction procedure begins. First, the occupying key is evicted and hashed to the location specified by its secondary function, where its contents are probed as before. This \textit{eviction chain} continues until either the evicted key is successfully inserted or a maximum chain length is reached. If the eviction is successful, then the new key is finally inserted at its primary location (first probe). Numerous follow-up studies to this canonical approach have introduced cuckoo hashing approaches with more than two hash functions (probes) per key, a separate hash table for each hash function, and other optimizations for concurrent, mixed operations (e.g., simultaneous inserts and queries) on the GPU. These studies are surveyed in this subsection.

Alcantara et al.~\cite{Alcantara:2009} introduce a data-parallel, dynamic hashing technique based on perfect hashing and cuckoo hashing that supports both hash table construction and querying at real-time, interactive rates. The querying performance of this technique is compared against that of the perfect hashing technique of Lefebvre and Hoppe~\cite{Lefebvre:2006} and a standard data-parallel sort plus binary search approach. In this work, a two-phase hashing routine is conducted to insert and query elements, with the goal of maximizing shared-memory usage during cuckoo hashing. 

First, elements are hashed into bucket regions within the hash table, following the perfect hashing approach of Fredman et~al.~\cite{Fredman:1984}. The maximum occupancy of each bucket is the number of threads in a thread block (e.g., 512), such that the entire bucket can fit within shared memory. The hash function aims to coherently map elements into buckets such that:
\begin{tightItemize}
\item Each bucket, on average, maintains a load factor of 80\%, and 
\item Spatially-nearby elements are located within the same bucket, enabling coalescing of memory among threads during queries.
\end{tightItemize}
If more than 512 elements hash to a given bucket, then a new hash function is generated and this phase is repeated. Then, within each bucket, cuckoo hashing is performed to insert or query an element, using $i=3$ different hash functions $h_{i}$ (i.e., the multiple choices), each corresponding to a sub-table $T_{i}$. During construction, each element simultaneously hashes to its location $h_{1}$ in $T_{1}$, in a winner-takes-all fashion. If multiple threads hash to the same location, then the winning thread (i.e., the last thread to write) remains and the other threads proceed to hash into location $h_{2}$ in $T_{2}$. This continues for $T_{3}$, after which any remaining unplaced elements cycle back to the beginning and hash into $h_{1}$ in $T_{1}$ again. At this point, if a collision occurs at $h_{1}$, then the current residing element is evicted and added to the batch of unplaced elements. This cuckoo hashing procedure continues until all elements are successfully placed into a sub-table $T_{i}$ or a maximum number of cycles have occurred.

An observation of this construction routine is that restarts can occur during both phases if either a bucket overflows or the cuckoo hashing reaches the maximum number of cycles within a bucket. While this reconstruction may be viewed as a disadvantage of probing techniques in general, the authors maintain that the occurrence of these restarts are reasonable in practice and fast to compute on massively-parallel GPU architectures. Moreover, this technique makes extensive use of thread atomics to increment and check values in both global and shared memory. While only a fixed number of atomic operations are made each phase, they are still serialized and must handle varying levels of thread contention, both of which are known to degrade performance.       

After construction, a query operation is performed by hashing once into a bucket, and then making at most $d=3$ hashing probes to locate the element within one of the sub-tables $T_{i}$ of the bucket. Insertions and queries are all conducted in a data-parallel, SIMD fashion. Since each thread warp assigned to a bucket has its own dedicated block of shared memory, the probing and shuffling of elements in the cuckoo hashing can be performed faster locally, as opposed to accessing global memory. 

Experimental results for this technique reveal the following:
\begin{tightItemize}
\item For querying elements (voxels in a 3D grid) in a randomized order, this hashing approach outperforms the perfect hashing approach of Lefebvre et al.~\cite{Lefebvre:2006} and the data-parallel binary search of radix-sorted elements of Satish et al.~\cite{Satish:2009}, particularly above 5 million elements. After this point, the binary searches used in both methods do not scale and become time-prohibitive.
\item For querying in a sequential order, the data-parallel binary search demonstrates better performance than this hashing technique, due to more favorable thread branch divergence and memory coalescing among the sorted elements. 
\item Constructing the hash table of elements in this approach is comparably-fast to radix sorting the elements, with noticeable slowdowns due to more non-coalesced write operations. Moreover, for large numbers of insertions, both approaches are magnitudes faster than constructing the perfect spatial hash table of~\cite{Lefebvre:2006}, which is initially built on the CPU, rather than the GPU (onto which the table is copied for subsequent querying).
\end{tightItemize}

Alcantara et al.~\cite{Alcantara:2011} improved upon their original work~\cite{Alcantara:2009} by introducing a generalized parallel variant of cuckoo hashing that can vary in the number of hash functions, hash table size, and maximum length of a probe-and-eviction sequence. In~\cite{Alcantara:2009}, the authors hypothesized that parallel cuckoo hashing within GPU global memory would encounter performance bottlenecks due to the shuffling of elements each iteration and use of global synchronization primitives; thus, shared memory was used extensively in the two-level hashing scheme. However, in this follow-up work, a single-level hash table is constructed entirely in global memory and addressed directly with the cuckoo hash functions, without the first-level bucket hash. The cuckoo hashing dynamics remain the same, except that the probing and evicting of elements occurs over the entire global memory hash table, as opposed to the shared-memory buckets of the two-level approach. 

To construct a hash table of $N$ elements, approximately $N$ threads will operate in SIMD parallel fashion to place their elements into empty slots in the global table. A given thread block will not complete until all of its threads have successfully placed their elements; a smaller block size helps minimize the completion time, as the block will likely contain fewer threads with long eviction chains.

The construction (insertion) and query performance of the single-level hash approach is compared against that of Merril's radix sort plus binary search~\cite{Merrill:2010} and the authors' previous two-level cuckoo hashing approach. Experimental results reveal the following:
\begin{tightItemize}  
\item \textit{Insertions}. For large numbers of insertions, the radix sort~\cite{Merrill:2010} becomes increasingly faster than both hashing methods, with a much higher throughput of insertions-per-second. For the same size hash table, the single-level hash table is constructed significantly faster than the two-level table, due to shorter eviction chains on average, over all insertion input sizes (the two-level table uses a fixed $1.42N$ space, while the single-level table is variable-sized). Radix sort achieves an upper bound of 775 million memory accesses (read and write) per second, while the single-level hashing only attains 670 million accesses per second. This higher throughput by radix sort is due to its more-localized memory access patterns that enable excellent memory coalescing among threads sharing a memory-bound instruction (up to 70\% of the theoretical maximum pin bandwidth on the tested Nvidia GTX 480 GPU, versus 6\% of the single-level hashing).
\item \textit{Queries: Binary Search vs. Hashing}. For random, unordered queries, binary search probing of the radix sorted elements is much slower than cuckoo hash probing of the elements. This arises from uncoalesced global memory reads and branch divergence for many of the threads, which use the maximum $O(logN)$ probes. Both cuckoo hashing approaches lookup elements in a worst-case constant number of probes and, thus, perform significantly better than binary searching, despite these probes being largely uncoalesced.
\item \textit{Queries: Two-Level vs. Single-Level}. When all queried elements exist in the hash table, the single-level cuckoo hashing makes a smaller average number of probes per query than the two-level approach, leading to faster completion times. However, when a large percentage of the queried elements do not exist in the hash table, the two-level hashing needs fewer worst-case probes before declaring the element as not found. This is because the single-level hashing uses four hash functions, or probes, to lookup an element, whereas the two-level hashing only uses three functions. By setting the number of hash functions to three in the single-level hashing, the authors observe comparable querying performance between the two approaches.
\end{tightItemize}

A notable performance observation in this work is that only randomized queries are considered. The authors indicate, as a limitation of their work, that if the elements to be queried are instead ordered (sorted), then binary searching the radix-sorted elements should yield improved thread branch divergence and memory coalescing. This work has since been incorporated into the CUDPP library~\cite{cuddp} as a best-in-class GPU hash table data structure.

\subsubsection{Multi-level and Bucketized Hashing}

Bucketized cuckoo hash tables (BCHT) organize groups of entries into \textit{buckets} (or bins), inside which cuckoo hashing is applied~\cite{Erlingsson:2006}. Typically, a single allocated hash table is used and partitioned into bucket regions, each of which may be assigned to a single warp of threads. Thus, the size of each bucket is uniform and proportional to the number of threads in a warp (e.g., 32 threads), the size of the cache line in a warp (e.g., 128 bytes), or the size of the shared memory within the warp's streaming multiprocessor (e.g., less than 50 kilobytes).  

As presented in section~\ref{cuckoo-hashing}, the bi-level design of Alcantara et al.~\cite{Alcantara:2009} performs bucketized cuckoo hashing by first perfect hashing into buckets that are the size of a thread block's shared memory, and then conducting cuckoo hashing within each bucket. Moreover, as presented in section~\ref{linear-probing}, the work of Bordawekar~\cite{Bordawekar:2014} develops a multi-level, bounded linear probing scheme. Sections~\ref{linear-probing} and~\ref{cuckoo-hashing} contain additional details on these approaches.

Zhang et al.~\cite{Zhang:2015} design a modified variant of a BCHT for use in accelerating queries of a GPU-based, in-memory key-value (IMKV) store, whose values reside in host memory. Traditionally, a bucket is the size of a thread warp or block (e.g., 512 threads) and each thread is assigned to a separate insert or query operation, with varying probe sequence lengths. However, with tens of thousands of threads operating on different buckets (warps) simultaneously, L2 cache (e.g., 1.5MB) contention will be high and likely lead to frequent evictions, which will force threads to perform multiple memory transactions. This work addresses this issue by sizing a bucket to a \textit{processing unit} of threads, which is set as a multiple of the memory transaction (L2 cache line) size (e.g., 32 bytes); the number of threads is the transaction size (in bytes) divided by 4 bytes, assuming each thread accesses a 4 byte (32-bit) key. Thus, a memory transaction services the entire processing unit, enabling coalescing among the threads. In a query operation, a key is first hashed to a bucket and given a key \textit{signature}, after which each thread in the unit compares its assigned key signature in the bucket with the query key signature. Via a warp-wide \textit{ballot} vote primitive, all threads indicate whether they have a match or not. If unsuccessful, the query key is hashed, via its second cuckoo hash function, into an alternative bucket. The same processing unit then searches for this key as before, reporting failure if it is not found. Insert operations are performed via a modified, bucketized cuckoo hashing routine in which a new key is only added into an empty slot, instead of evicting a collided key.

Breslow et al.~\cite{Breslow:2016} introduce an expansion of a BCHT that allows for higher load factors, improved bucket load balancing, and a lower expected number of bucket lookups (less than 2) for both positive and negative queries. In this \textit{Horton table}, a row is maintained for each bucket, which is denoted as either \textit{Type A} or \textit{Type B}. Each key is hashed by its \textit{primary hash function} into the \textit{primary bucket}. If the primary bucket is full, then the key either hashes, via one of its \textit{secondary hash functions}, to a \textit{secondary bucket}---after which we denote the key as a \textit{secondary item}---or replaces a secondary item in the primary bucket. If the key is a secondary item, then it is placed in the secondary bucket that is least full; note that several secondary hash functions (and buckets/rows) can be specified. Then, the filled primary bucket is \textit{promoted} (if not already) to Type B and its last stored key is evicted (moved to a secondary bucket) to make room for a compact \textit{remap entry array} that stores an index, or \textit{remap entry}, to the secondary bucket of each secondary item. This important feature permits all secondary items to be efficiently tracked, allowing no more than two probes and hash function evaluations per query. Additional bookkeeping and logic is used to delete keys and handle a cascading effect where an evicted key causes its secondary bucket to convert into Type B, which induces another eviction, and so on.

Experimental results of large query sets reveal that most successful lookups occur within the primary buckets, allowing a high load factor with only one hashing probe. Moreover, the performance of the Horton table is compared against that of a baseline BCHT similar to Zhang et al.~\cite{Zhang:2015}. For all successful queries, the Horton table increases throughput (billions of keys queried per second) over the baseline by 17 to 35 percent. For a set of all unsuccessful queries, the Horton table increases throughput by 73 to 89 percent over the baseline, needing only one hash probe to detect failure. An important note regarding the design and performance of this approach is that only the query operations are conducted in data-parallel fashion on the GPU. The detailed insertion and construction phase is run on the CPU, which could make reconstruction costly for use other than as a static hash table, which is sufficient for the query-heavy usage of most key-value store systems~\cite{Zhang:2015}.

Hetherington et al.~\cite{Hetherington:2015} develop a fixed-sized \textit{set-associative} hash table for scaling-up the throughput of key-value storage and accesses. As part of a \textit{MemcachedGPU} caching service, HTTP GET requests are parsed to extract query keys that are hashed to a hash table on the GPU. This table facilitates $k$-way, set-associative hashing with each set (or bucket) entry consisting of an 8-bit key identifier hash and a pointer to the actual memory address (within a dynamically allocated \textit{slab} of memory) at which the key is stored. After hashing to a set, each query key compares itself to the 8-bit hash and, if a positive match, accesses the key in memory and instigates a return package with the associated value, which is stored in CPU memory. If the query key does not exist in the set, then it was previously a \textit{least-recently-used} (LRU) key that must have been evicted from the set by a colliding, more-recent key in the same set (recent usage based on timestamp). Thus, an HTTP SET request can reinsert this key into an empty entry in the set or evict a LRU key that resides in the set. Each set maintains and updates its own local LRU array. Experiments over varying hash table sizes (number of entries) and a query-heavy distribution of key-value requests ($95\%$ GET and $5\%$ SET) reveal that MemcachedGPU achieves an acceptable hash table miss rate with $16$-way set associativity. In these experiments, each request key is assigned to an individual warp thread for SIMT execution. Unless requests exhibit spatial locality, branch and memory divergence are inevitable in this approach.       

Ashkiani et al.~\cite{Ashkiani:2016} design a set of \textit{multisplit} data-parallel primitives for the GPU that efficiently permute elements into contiguous buckets. While this study is not focused on hashing, it recommends that the multisplit can be used to map elements into the first level of buckets in a multi-level hash table, such as the two-phase hash table of Alcantara et al.~\cite{Alcantara:2009}. Moreover, this work contributes a \textit{reduced-bit} radix sort that converges to and exceeds the performance of state-of-the-art radix sort~\cite{Merrill:2010} as the number of buckets is increased. Thus, if the order of insertions and queries into a bucket-based hash table are non-random and ordered, then this sorting primitive may offer an effective substitution for a bucketing procedure. These primitives have since been incorporated into the CUDPP library~\cite{cuddp}.

\subsubsection{Double Hashing}
\label{double-hashing}

\textit{Double hashing} first hashes a key $k$ to location $h(k)$ in the hash table and then, if the location is already occupied, computes another independent hash $h'(k)$ that defines the step size to the next probing location~\cite{Cormen:2001}. Thus, the second probe location is $h(k) + i\cdot h'(k)$, where $i$ is the current $i$-th probe in the probe sequence. This hashing and probing continues until an empty slot (insertion) or $k$ itself (query) is found. Similar to linear and quadratic probing, if $h(k)$ is empty, then $k$ is inserted immediately, without probing. The choice of the second hash function has a large impact on performance, as it dictates the locality of consecutive probes and, thus, the opportunity for memory coalescing among threads on the GPU.

Khorasani et al.~\cite{Khorasani:2015} introduce a \textit{stadium hashing} (\textit{Stash}) technique that builds and stores the hash table in out-of-core host memory, and resolves insert collisions via double hashing until an empty slot is found. In GPU global memory, a compact auxiliary \textit{ticket-board} data structure is maintained to grant read and write accesses to the hash table. For each hash table location, the ticket board maintains a \textit{ticket}, which consists of a single \textit{availability bit} and small number of optional \textit{info bits}. The availability bit indicates whether the location is empty (set to 1) or occupied by a key (set to 0), while the info bits are a small generated signature of the key to help identify the key prior to accessing its value. Within individual thread warps, a shared-memory, \textit{collaborative lanes} (\textit{clStash}) load-balancing scheme is used to ensure that, during insertions, all threads are kept busy, preventing starvation by unsuccessful threads.   

Stadium hashing is meant to address three limitations of previous GPU parallel hashing techniques, specifically in regard to the cuckoo hashing approach of Alcantara et al.~\cite{Alcantara:2011}:
\begin{tightEnumerate}
\item Support for concurrent, mixed insert and query operations. Without proper synchronization, cuckoo hashing encounters a race condition whereby a query probe fails at a location because a concurrently-inserted key hashes to the location and evicts the queried key, yielding a false negative lookup. Stadium hashing avoids this issue by using eviction-free double hashing and granting atomic access to a location via a ticket board ticket with the availability bit set to 1.
\item Reduce host-to-device memory requests for large hash table sizes. In cuckoo hashing, CAS atomics are used to retrieve the content of a memory location, compare the content with a given value, and swap in a new value, if necessary. When a hash table is stored in host memory, the large number of parallel retrieval requests from thousands of GPU threads will turn the hashing into a PCIe bandwidth-bound problem and degrade performance. Stadium hashing uses the GPU ticket board data structure to minimize retrieval requests to the host memory hash table.
\item Efficient use of SIMD hardware. During a cuckoo hashing operation, a thread failing to insert or query a key can cause starvation among the other threads in the thread warp, as they all perform the same instruction in lock-step. Thus, if the other threads complete their operation early, then they will remain idle and contribute to work imbalance. Stadium hashing uses the clStash load-balancing routine to maintain a warp-wide, shared memory store of pending operations that early-completing threads can claim to remain busy. 
\end{tightEnumerate}

For an out-of-core hash table, the ticket-board with larger ticket sizes (more info bits per key) helps improve the number of operations per second by reducing the number of expensive host memory accesses over the PCIe bus. This improvement is especially significant for unnecessary queries of elements which do not actually reside in the host hash table. In this case, the PCIe bandwidth from GPU to CPU memory is the primary performance bottleneck. However, when the hash table resides in GPU memory, the underutilization of SIMD thread warps becomes the primary bottleneck on performance for low load factors (fewer collisions). The efficiency of warps is shown to improve by using the collaborative lanes clStash scheme in combination with the Stash hashing.

Regarding the experiments and findings in this work, the cuckoo hashing approach of ~\cite{Alcantara:2011} is specifically designed for hash table construction and querying within GPU memory. Thus, the use of this technique in out-of-core memory should not necessarily be expected to perform optimally, and should be kept in mind when comparing against the out-of-core performance of stadium hashing. 

\subsubsection{Robin Hood-based Hashing}
\label{robin-hood-hashing}

\textit{Robin Hood} hashing~\cite{Celis:1986} is an open-addressing technique that resolves hash collisions based on the \textit{age} of the collided keys. The age of a key is the length of the probe sequence, $h^{1}(k), h^{2}(k),\dots$, needed to insert the key into an empty slot in the hash table. During a collision at a probe location, the key with the youngest age is evicted and the older key inserted into that location. The evicted key is then robin hood hashed again until it is placed in a new empty location, incrementing its age along the new probe sequence. The idea of this approach is to prevent long probe sequences by favoring keys that are difficult to place. Even in a full table with high load factor, this eviction policy guarantees an expected maximum age of $\Theta(\log n)$ for an insert or query key. However, the worst-case maximum age $M$ may still be higher and worse than the maximum probe sequence length of cuckoo hashing, prompting a table reconstruction in some cases. These maximum $M$ probes will be required during queries for empty keys (those which do not exist in the hash table), unless they are detected and rejected early.      

Garcia et al.~\cite{Garcia:2011} introduce a data-parallel robin hood hashing scheme that maintains coherency among thread memory accesses in the hash table. Neighboring threads in a warp are assigned neighboring keys to insert or query from a spatial domain (e.g., pixels in an image or voxels in a volume). By specifying a coherent hash function, both keys will be hashed near each other in the hash table and the threads can access memory in a coalesced fashion, i.e., as part of the same memory transaction. Thus, the sequence of probes for groups of threads will likely also be conducted in a coherent manner, as nearby keys of a young age are evicted and replaced by nearby keys of an older age. 

The insertion and query performance of this techniqie is evaluated on both randomly- and spatially-ordered key sets from a 2D image and 3D volume. For all load factor settings, the existence of coherence in the keys and probe sequences results in significant improvements in construction and querying performance (millions of keys processed per second), as compared to the use of randomly-ordered keys. Moreover, coherent hashing achieves greater throughput than the cuckoo hashing of Alcantara et al.~\cite{Alcantara:2011}, which employs four hash functions for a maximum probe sequence of length four. For load factors above $0.7$, coherent hashing maintains superior performance without failure (hash table reconstruction) during insertions, whereas the cuckoo hashing exhibits an increase in failures. 

In absence of coherence in the access patterns, coherent hashing brings little to no benefit compared to the random access robin hood and cuckoo hashing. Thus, this approach is of particular use for applications with spatial coherence in the data. In one of the spatially-coherent experiments, the task is to insert a sparse subset of pixels from an image (e.g., all the non-white pixels) into the hash table, and then query every pixel to reconstruct the image. Since only non-white pixels are hashed, there will be empty queries for the white pixel keys. Spatial and coherent ordering of keys is attained by applying either a linear row-major, Morton, or bit-reversal function to the spatial location of elements; a non-coherent, randomized order is attained by shuffling the keys. 

Coherent hashing has some notable design characteristics that can affect GPU performance. First, upon completing an insert or query operation, a thread sits idle until all threads in its warp have finished as well. The number of threads per warp (192 in this work) and amount of branch divergence due to incoherent ordering of keys are primary factors affecting the warp load balancing. Second, while inserting a key, the hash table is reconstructed if the age, or probe sequence length, of the key exceeds a threshold maximum (15 in this work). Moreover, the hash table is not fully dynamic and is designed to process queries after an initial build phase. Thus, if new keys are inserted after the build phase, then the table is reconstructed entirely, with a larger table size or load factor if necessary. 

Zhou et al.~\cite{Zhou:2016} design a modified GPU-based robin hood hashing scheme for use in storing and extracting the top-$k$ most similar matches, or results, of a query (e.g. a document of words or multi-dimensional attribute vector). In this similiarity search, a two-level Count Priority Queue (c-PQ) data structure hashes potential candidates for the top-$k$ results into an upper-level hash table, as determined by a lower-level \textit{ZipperArray} histogram array of object counts, where $ZipperArray[i]$ is the number of objects (e.g., word or attribute value) that appear $i$ times in the query (count of $i$). An \textit{AuditThreshold} is set as the minimum index (count) of $ZipperArray$ whose value (number of objects) is less than $k$. For an object to be inserted into the hash table, it must have a count greater than or equal to the \textit{AuditThreshold}. As new items are added, objects counts and the $ZipperArray$ are updated, and the \textit{AuditThreshold} may be increased. During insertion into the hash table, the robin hood hashing scheme of Garcia et al.~\cite{Garcia:2011} is used, with an additional feature that an object with a count smaller than $(AuditThreshold - 1)$ can be directly overwritten, and not evicted, during a hash collision. This helps reduce the average age, or probe length, of new insertions, as previously-inserted objects become \textit{expired} due to an increase in the $AuditThreshold$. This modification, along with a lock-free synchronization mechanism, affectively contributes a dynamic hash table variant of ~\cite{Garcia:2011}. 
%\fix{Sam: Is it possible to provide an overall assessment/evaluation/conclusion of the 5 techniques from the ``open-address probing'' class? Since section 3.1 is a pretty long one, an overall assessment would be helpful for the readers.}

\subsection{Perfect Hashing}
\label{perfect-hashing}

Whereas the previous hashing categories employ \textit{imperfect} hash functions that require collision-handling and multiple probes, \textit{perfect hashing} maps each key to a unique address in the hash table, resulting in no collisions and enabling single-probe queries. If the length of the hash table $m$ is equal to the number of keys $n$, then a perfect hash function over the keys is \textit{minimal} and effectively scatters, or permutes, the keys within the table.

In theory, obtaining a perfect hash function, especially for large sets of keys, is a difficult, low-probability task. Given a universe $U = \{0,1,\dotsc,u-1\}$ of possible keys, subset $S \subset U$ of $|S| = n$ keys to be hashed, a hash table of size $m$, and class $H$ of hash functions $h: U \to [0,m-1]$, the probability $P(n,m)$ of randomly placing $n$ keys in $m\geq n$ slots without a collision is
\begin{align*}
P(n,m) &= \frac{m\left(m-1\right)\cdots \left(m-n+1\right)}{m^n}.
\end{align*}
This can also be stated as the probability of a randomly-chosen hash function $h \in H$ being a perfect hash function over the set $S$. As a reinterpretation of the classical birthday paradox, only one in ten million hash functions $h \in H$ is a perfect hash function for $n=31$ keys mapped into $m=41$ locations. When $m \gg n$, $P(n,m) \approx e^{\frac{-n^2}{2m}}$, which implies that there is a very low probability of attaining a perfect hash when the load factor or occupancy of the hash table is very high; i.e., $m \ll n^2$. Moreover, when $m=n$, $P(n,m)=\frac{n!}{n^n}$, which is the probability of achieving a minimal perfect hash~\cite{Mehlhorn:1982}. For larger key set sizes $n$, such as those seen in practical applications, the minimal perfect hash probability decreases very rapidly and is approximated as $e^{-n}$.

In practice, a perfect hash function can be described as an imperfect hash function that is then iteratively corrected into a perfect form. One approach to doing this is to construct one or more auxiliary lookup tables that perturb the hash table addresses of collided keys into non-colliding addresses~\cite{Fox:1992}. These tables are typically significantly more compact than the hash table. Another foundational approach, introduced by Fredman et al.~\cite{Fredman:1984}, is the use of a multi-level hash table that hashes keys into smaller and smaller buckets---each with a separate hash function---until each key is addressed to a bucket of its own, yielding a collision-free, perfect hash table with constant worst-case lookup time. Moroever, a perfect hash function is most suitable for static hash tables (and key sets), in which no insertions or deletions occur after the construction of the table. If dynamic updates are performed, then the function will inevitably become imperfect---with collisions among relocated keys---and require a reconstruction procedure. Czech et al.~\cite{Czech:1997} survey a rich body of related work investigating additional perfect hashing and minimal perfect hashing schemes (largely non-parallel), each designed for CPU-based storage and computation.    

Lefebvre and Hoppe\cite{Lefebvre:2006} introduce a perfect spatial hashing (PSH) approach that is also the first GPU-specific perfect hashing approach. In PSH, a minimal perfect hash function and table are constructed over a sparse set of multi-dimensional spatial data, while simultaneously ensuring locality of reference and coherence among hashed points. Thus, on the GPU, spatially-close points are queried coherently, in parallel, by threads within the same warp. In order to maximize memory coalescing among these threads, points are also coherently accessed within the hash table, as opposed to via a random access pattern, which can necessitate multiple memory load instructions.

In the PSH study, the spatial domain $U$ is a $d$-dimensional grid with $u$ points (positions), where $d \in \{2,3\}$. The grid serves as a bounding box over a sparse subset $S \subset U$ of $n$ points that have associated data records $D(p),p \in S$ (e.g., RGB color value for each pixel or voxel); thus, the remaining $u-n$ grid locations are stored in memory without any compute value. The sparse subset $D$ is more-compactly stored in a dense hash table $H$ of size $m \geq n$. This table is addressed by a multi-dimensional perfect hash function $h$ that is composed of two imperfect hash functions, $h_0$ and $h_1$, and an \textit{offset table} $\Phi$ that ``jitters'' $h_0$ into perfect form. This function is defined as \[ h(p) = h_0(p) + \Phi [h_1(p)]\bmod \bar{m}, \forall p \in S, \]
where $h_0$ and $h_1$ perform simple modulo addressing and wrap the points $S$ multiple times over $H$ and $\Phi$, respectively. Due to this modulo, lockstep addressing by $h_1$, spatial coherence is preserved for accessess into $\Phi$. However, the values $\Phi[h_1(p)]$ perturb the coherency of the combined function $h$. By constructing $\Phi$ such that adjacent offset values are locally constant, the coherency of $h$ can be ensured. Note that $h$ is not strictly a minimal perfect hash function (i.e., when $m=n$), since the hash table size $m$ may need to be slightly increased to faciliate the perfect hash represented within $\Phi$. The number of unused table entries $m-n$ is kept small and is considered \textit{near-minimal}. 
Thus, the sizing of $H$ and the spatial coherency of $h$ are both dependent on the proper construction of $\Phi$.

The construction of the offset table $\Phi$ proceeds by first identifying the smallest table size $r$ that yields a perfect hash $h$. A geometric progression or binary search of $r$ is iteratively conducted, depending on whether faster construction or a compact representation is desired, respectively. For each size $r$ tested, the offset values $\Phi[q], 0 \leq q \leq r$ are assigned via a greedy heuristic procedure that seeks to maximize spatial coherence. Since $r < n$, on average $\lfloor \frac{n}{r} \rfloor$ points, $h_{1}^{-1}(q) \subset S$, hash to each entry $q \in \Phi$. The entries $q$ are assigned in descending order by size of their point sets $h_{1}^{-1}(q)$. Then, each $q$ is assigned one of the following two candidate heuristic values:
\begin{tightEnumerate}
\item Same offset value as a neighboring entry, $\Phi[q'],\|q-q'\| < 2$.
\item For each $p \in h_{1}^{-1}(q)$ with a neighboring point $p' \in S, \|p-p'\| = 1$ already hashed in $H[h(p')]$, the offset value that places $p$ in an empty neighboring slot of $H[h(p')]$.
\end{tightEnumerate}
If $\Phi$ yields a perfect hash function $h$ with the tested size $r$, then the construction phase completes; otherwise, the routine is conducted again with another size $r$. 

In a SIMT fashion on the GPU, point queries $q \in U$ are executed in parallel by threads, each computing $h(q)$ and looking up an associated value from $H[h(q)]$, which is mapped to a single point due to the perfect hash $h$. If $q$ does not exist in $H$, then a negative query result is returned. 

Note that the construction of $\Phi$ is an inherently sequential process, since the assignment of offset values depends on earlier offset values or hashed points in $H$. Moreover, the construction of $H$ and $\Phi$ is performed on the CPU in this work, due to the larger memory requirements and presumed usage as a pre-processing step; thus $H$ must be copied into GPU device memory over the PCIe bus. Moreover, $H$ is designed to be static, since insertions or deletions of points after construction destroy the perfect hash and require $\Phi$ to be reconstructed.

Bastos and Celes~\cite{Bastos:2008} implement a GPU-based \textit{link-less} octree data structure by hashing the parent-child (node) relationships into the perfect hash table of Lefebvre et al.~\cite{Lefebvre:2006}. Thus, instead of constructing a multi-level tree with pointers over sparse spatial data, only a compact perfect hash table needs to be built; however, updates to this data structure are costly and require the entire table to be reconstructed, as the perfect hash is intricately data-dependent. Since perfect hashing is collision-free, direct random-access queries can be made on the octree, as opposed to traditional pointer tracing in tree traversals. 

Choi et al.~\cite{Choi:2009} follow-up the work of Bastos and Celes~\cite{Bastos:2008} with a similar link-less octree design that avoids the need to store extra bitmaps for the \textit{sparsity encoding} of empty grid cells in the sparse spatial domain. This encoding indicates whether a cell contains associated data that is stored within the hash table; if no data is present, then a query operation for the cell can be avoided. While this latter approach significantly reduces storage costs, it executes random-access queries much slower than similar accesses into the benchmark pointer-based octree. 

Schneider and Rautek~\cite{Schneider:2017} denote sparsity encoding as a memory overhead cost for providing \textit{unconstrained access}, or empty cell querying, in the spatial perfect hashing approach of Lefebvre et al.~\cite{Lefebvre:2006}. This study proposes a compact, GPU-based \textit{Fenwick tree} data structure that supports unconstrained accesses without additional occupancy storage to denote empty cells.

\subsection{Spatial Hashing}
\label{spatial-hashing}

The following two subsections present GPU-based spatial hashing techniques for inserting and querying regular grid cells (subsection~\ref{sec:grid-based}) and point coordinates (subsection~\ref{sec:locality}) within a multi-dimensional spatial domain.

%kd-tree: Recursively divide space in half
%binary spatial partitioning: splitting planes can be arbitrarily located, non-AABB because of non-aligned planes
%Recursion is stopped when a bin contains fewer than a given number of points or recursion has reached a maximum number of subdivisions
%Organize objects into a tree
%Each node in the tree contains an axis-aligned bounding box (AABB)
%bounding box of all the objects below it
%Group objects in the tree based on
%spatial relationships
%hierarchical grouping of 3D objects, that subdivides objects into smaller and smaller groups until each object is contained within its own group
%The idea is to construct a uniform 3D grid whose cells are at least the same size as the largest object.
%The grid cell size, which is used for spatial hashing, influences the number of object primitives, that are mapped to the same hash index

\subsubsection{Grid-based Spatial Hashing}
\label{sec:grid-based}
Most real-world use cases of searching require a data structure that can support lookups of geometric primitives --- e.g., point coordinates, polygonal shapes, and voxels --- that exist within a multi-dimensional \textit{spatial domain}, such as $\mathbb{R}^2$,~$\mathbb{R}^3$, or $\mathbb{R}^n$. One approach is to explicitly compute a bounding box over the domain and then recursively subdivide it into smaller and smaller regions, or cells, which contain a group of primitives or a subset of the spatial domain. This subdivision hierarchy can be represented by a grid (e.g., uniform and two-level) or tree (e.g., $k$-d tree, octree, or bounding volume hierarchy) data structure (see Subsection~\ref{gpu-data-structures}) that conducts a query operation by traversing a path through the hierarchy until the queried primitive is found. While these structures are designed for fast, highly-parallel usage, they typically do not exhibit fast reconstruction rates due to complex spatial hierarchies, and may contain deep tree structures that are conducive to thread branch divergence during parallel query traversals. These attributes are particularly important to real-time, interactive applications, such as surface reconstruction and rendering, that make frequent updates and queries to the acceleration structure.

An alternative approach that addresses these limitations is to perform \textit{spatial hashing} over the primitives, whereby the multi-dimensional domain is projected, or compressed, to a single dimension in the form of a hash table data structure. Instead of computing a bounding box over the spatial domain and explicitly storing the entire space, spatial hashing assumes an implicit, infinite regular grid over the domain and maps each positional primitive (e.g., a point coordinate) to a uniformly-sized and axis-aligned cell within the grid. Each cell is uniquely addressed by unit coordinates and contains a user-specified number of primitives within its bounds~\cite{VTK:Book}. These coordinates are used by the hash function to hash the cell into the hash table. Two or more cells may hash to the same address, resulting in collisions that must be resolved. To query a primitive (or cell), the primitive is mapped to its cell and the cell is hashed to an address in the hash table. From this address, the cell is searched, using more than one probe if a collision occurs. Typically, to exploit sparsity, only non-empty cells that contain computable primitive data (e.g., pixel intensity, RGB, or density) are inserted into the hash table. A query of an empty cell will return a negative result, as it doesn't exist in the table. 

This canonical grid-based \textit{voxel hashing} approach was introduced by~\cite{Teschner:2003} as a CPU-based search structure for detecting colliding 3D tetrahedral cells in $\mathbb{R}^3$ domain space. Several follow-up studies have since introduced GPU-based spatial hashing techniques based off of this approach, and they are surveyed as follows.

%Each cell in the structured grid can be addressed by index (i, j) in two dimensions or (i, j, k) in three dimensions, and each vertex has coordinates $(i\cdot dx,j\cdot dy) (i\cdot dx,j\cdot dy)$ in 2D or $(i\cdot dx,j\cdot dy,k\cdot dz) (i\cdot dx,j\cdot dy,k\cdot dz)$ in 3D for some real numbers dx, dy, and dz representing the grid spacing.

%As previously covered in section~\ref{perfect-hashing}, Lefebvre and Hoppe~\cite{Lefebvre:2006} design a GPU-based hashing scheme that couples both perfect hashing and spatial hashing, and achieves significant compaction of sparse 2D images and 3D surfaces, which are traditionally stored as grids that contain many data-less (or empty) grid positions. Moreover, the robin hood hashing of Garcia et al.~\cite{Garcia:2011} is designed for coherent accesses in spatial domains, such as images and voxel grids (see section~\ref{robin-hood-hashing}).

Nie{\ss}ner et al.~\cite{Niener:2013} extend the approach of Teschner et al.~\cite{Teschner:2003} with more sophisticated collision handling and a 3D voxel hashing scheme that is designed particularly for fast, real-time hash table updates on the GPU. An infinite uniform grid subdivides the world space into voxel blocks, each of which consists of $8^3$ voxels. The world coordinates of each voxel block are hashed as an address into a bucketed hash table. During an insertion, a block probes linearly through its assigned bucket for the first empty slot that it can occupy. If a free slot is found, then the block is inserted. Otherwise, if the bucket is already full, then \textit{overflow} occurs and a linked list entry in the last slot points to the next free slot in another bucket of the hash table. The block then probes along this overflow chain to find the next empty slot. Due to interconnection among buckets, each hash entry contains an offset pointer to its neighboring bucket entry, which may not be adjacent in the table. A query operation conducts similar probing to find a particular block within the hash table. Additionally, lighweight GPU atomic primitives are used to coordinate data-parallel insertions and deletions of blocks, each assigned to an individual thread. While an entire bucket is locked for writing during an insertion into the bucket, no degradation in performance is observed for a high-throughput, real-time 3D scene reconstruction experiment. Moreover, by using a larger hash table size, the number of collision is kept minimal and prevents bucket overflows into other disparate buckets, which can cause uncoalesced memory accesses among warp threads.

K{\"a}hler et al.~\cite{Kahler:2016} introduce a GPU-based hierarchical voxel block hashing technique that uses multiple hash tables in a hierarchy to store finer and finer resolutions of grid discretitzation for voxel blocks (cells). Initially, each block is hashed to an entry in a first-level hash table of \textit{coarse resolution}. Then, if the voxels within this block are represented at a \textit{finer resolution}---as indicated by a flag in the entry of each hash entry---the block is hashed again with a different hash function into a second-level hash table. This hierarchical hashing continues until an entry is reached that contains a pointer to the \textit{voxel block array}, which stores the actual, individual block data. Atomic voxel block \textit{splitting} and \textit{merging} operations are supported to enable the addition or removal of hash table entries for finer or coarser resolutions, respectively. On scene reconstruction experiments with signed distance function (SDF) values for roughly 20 million points, this adaptive hierarchical representation, with 3 resolution levels and base voxel size of 2 mm, attains greater accuracy than a fixed representation with 8 mm voxel sizes.   

Note that the hierarchical voxel hashing of K{\"a}hler et al.~\cite{Kahler:2016} is inspired by the general approaches of Eitz and Lixu~\cite{Eitz:2007} and Pouchol et al.~\cite{Pouchol:2009}, which themselves expanded upon the original spatial hashing of Teschner et al.~\cite{Teschner:2003}. These studies are each CPU-based and use real-time collision detection as a motivating application.

Chentanez et al.~\cite{Chentanez:2016} introduce a GPU-based spatial hashing variant of Teschner et al.~\cite{Teschner:2003} for detecting and deleting overlapping triangles on the surface of a 3D mesh volume, as vertices are advected (i.e., mesh refinement). In this work, the 3D bounding cells of triangles are inserted into a specially-arranged hash table using the coordinate-based hash function from~\cite{Teschner:2003}. The hash table consists of $n$ buckets each with $m$ available slots ($n\cdot m$ entries), and the first $n$ entries of the table are reserved to store counts of the number of slots $j \leq m$ that are occupied in each bucket. Thus, the total allocated size of the table is $n(1+m)$. During an insertion of a cell $k$ into bucket $h(k)=b$, the thread assigned to cell $k$ first checks the occupancy count value for bucket $b$. If $b$ has open slots, then $k$ is inserted into the first available slot and the count for $b$ is atomically incremented. Otherwise, the thread examines the count for the next bucket $b+1$ and inserts $k$ into the first open slot of $b+1$, if possible, so on and so forth until $k$ is successfully inserted. This is a modified collision resolution scheme whereby a bucket collision only occurs when the bucket is full and subsequent buckets are then linearly-probed for one that has an empty slot. During a cell query, the same linear probing over buckets is performed, beginning with the bucket to which the cell is hashed. 

Note that, in this approach, thousands of other parallel threads are executing the same operation on different triangle cells, likely resulting in high contention for atomic writes for the bucket count values and worst-case linear probing sequences that induce branch divergence within warps. The extent of such divergence depends on the size $m$ of each bucket and whether locality of reference is maintained among bucket entries when hashing spatially-nearby cells. These performance effects are not explored in the experimental findings of this approach.     

Tumblin et al.~\cite{Tumblin:2015} expand upon traditional perfect spatial hashing (PSH) with a \textit{compact spatial hashing} (CSH) variant that compacts a perfect hash table when it becomes too sparse, saving unused memory on the GPU. As a larger number of keys need to be hashed, a sufficiently large hash table must be allocated to construct a perfect hash among the keys. Often, this large table still contains many empty locations, resulting in a low occupancy and high \textit{compressiblity}, which is the ratio of available table locations to occupied locations. A \textit{compression} function randomizes the original hash locations of each key and fits them within a smaller, compact hash table of size proportional to the number of keys divided by a desired load factor. Since PSH is collision-free, this compaction inevitably induces collisions, which are handled in this work by a canonical quadratic probing open-addressing method in parallel. The goal of the compression function, thus, is to reduce the occurrence of collisions via random scattering of keys. However, this random re-assignment of hash locations disrupts any spatial locality that existed among the keys, preventing warp-level memory coalescing during accesses. Experimental results for an adaptive mesh refinement (AMR) task show that as the perfect hash table reaches 20 to 40 times the size of the compact hash table, the CSH becomes the faster method. Thus, the exceedingly larger memory of PSH offsets the extra costs (e.g., thread divergence and uncoalesced memory) of resolving collisions and querying in CSH.  

Duan et al.~\cite{Duan:2017} present an \textit{exclusive grouped spatial hashing} (EGSH) scheme that is optimized to compactly represent multi-dimensional domains that contain repetitive data (e.g., duplicate RGB or density values). The goal of this approach is to identify all groups of points that share the same data values and then, for each group, compress its points into a single group-wide value, avoiding the unnecessary storage of duplicates, which are significantly prevalent in some domains. This grouped hashing is performed over multiple iterations using multi-level hash tables until each group has been exclusively hashed into a unique table location. The logistics of this approach are discussed as follows:
\begin{tightItemize}
\item In the first iteration, all points in the \textit{input domain} are hashed into a hash table of size equal to the number of points. Then, at each non-empty hash table location, collided points are binned into disjoint groups based on their corresponding data values. The data value of the group with the most points is set as the exclusive value in this hash table location, replacing and compressing the many repetitive points of the group. For subsequent querying, the hash table ID (iteration index) of each compressed point is stored in a global, persistent \textit{coverage table}. The remaining uncompressed points of the other groups are then advanced as the input domain to another iteration of exclusive group hashing. In the next iteration, all the uncompressed points (among all hash table locations from the previous iteration) are hashed into a smaller hash table of size approximately equal to the number of groups from the previous iteration. The grouping and compression routine is then conducted as before, and the hashing iteratively continues until all points are compressed.
\item The compression of repetitive elements contrasts with other hashing approaches covered in this survey, which hash repetitive keys into separate, and possibly disparate (depending on load factor and table size) addresses of the table upon collision.
\end{tightItemize}

Experiments on the GPU reveal that after several iterations of EGSH, the input domain becomes very sparse and has a rapid reduction in the amount of repetitive data (uncompressed groups). Both of these traits are highly suitable for the perfect spatial hashing (PSH) of Lefebvre and Hoppe~\cite{Lefebvre:2006}, which similarily provides constant-time random accesses. Thus, an optimized variant of EGSH performs exclusive grouped hashing for a small number, $k$, of iterations---generating $k$ levels of hash tables---and then applies the PSH on the remaining uncompressed input domain. In this work, $k=6$ is used for a set of 2D and 3D grid-based input textures, all of which possess a high ratio of points with repetitive data values.
The results of a comparison between optimized EGSH and PSH on these textures reveal that both schemes have similar constant access times, while the EGSH memory cost is typically less than half of the PSH memory cost. Takeaways of this study are that PSH achieves best performance on sparse, slightly repetitive domains, as opposed to sparse, highly repetitive or dense domains. Contrarily, EGSH attains the smallest memory savings and construction time for input domains with highly repetitive data.

A few important notes regarding this EGSH work are the following:
\begin{tightItemize}
\item Thread- and warp-level GPU performance findings are not provided. Only high-level runtimes and memory usage are analyzed. Moreover, the ESGH multi-level hash tables appear to be constructed on the CPU and copied over the PCIe bus to the GPU for subsequent querying, much like that of the PSH hash table construction. This is notable, since, during construction, the search for the group with the maximal number of elements at each hash table location is the most time-consuming task and may not be optimally parallelized on the CPU.
\item When each group has only one point, ESGH degenerates into PSH, whereby all points are hashed to unique locations of a single table. When the groups contain multiple repetitive points, multiple sub-tables are needed to complete the ESGH.
\end{tightItemize}

\subsubsection{Locality-Sensitive Hashing}
\label{sec:locality}

Much like grid-based spatial hashing, locality-sensitive hashing (LSH) reduces the dimensionality of high-dimensional data via a projection to a 1D hash table. LSH hashes input elements so that similar elements map to the same buckets with high probability, with the number of buckets in the hash table being much smaller than the universe of possible input elements. LSH differs from the other hashing approaches covered in this survey because it aims to maximize the probability of collisions between similiar items~\cite{Indyk:1998}. Similar to other approaches, LSH employs a collision resolution scheme to relocate collided elements that are inserted into the hash table. During a query operation, LSH is well-suited for returning the $k$ approximate nearest neighbors ($k$ANN) of the query element $q$, since these neighbors will likely reside in the same bucket as $q$~\cite{Indyk:1998,Datar:2004}. While performant GPU-based, brute-force approaches exist to find the \textit{exact} $k$ nearest neighbors~\cite{Kuang:2009,Li:2015}, a large body of recent research has investigated the design and performance of LSH for $k$ANN.

More formally, canonical LSH proceeds as follows, beginning with the construction of the LSH hash tables. Given a set of $D$-dimensional points $S \subset \mathbb{R}^D$,~ $M < D$ different hash functions $h_{i}(\mathbf{p}) = \lfloor (\mathbf{a}_{i} \cdot \mathbf{p} + b_i)/W \rfloor$ are used to project each point $\mathbf{p} \subset S$ to a cell within a $\mathbb{Z}^M$ lattice space, the size of which is determined by $M$ and $W$ ($\mathbf{a}_i \in \mathbb{R}^D$ and $b_i \in [0,W)$ are randomly generated). This cell location, or \textit{LSH projection}, is specified as a $M$-dimensional vector $H_{j}(\mathbf{p})= \langle h_1(\mathbf{p}),h_2(\mathbf{p}),\cdots,h_M(\mathbf{p})\rangle$ and then mapped into a single value $g_j(H_j(\mathbf{p}))$ known as the \textit{LSH code}. This code represents the bucket index of the hash table $G_{j}$ into which $\mathbf{p}$ is then inserted. To decrease the collision probability of false neighbors, $\mathbf{p}$ is projected and hashed into $j = L$ different and independent hash tables, with each instance of $H_{j}$ and $G_{j}$ being randomly generated. During a query for arbitrary point $\mathbf{v} \in \mathbb{R}^D$, $\mathbf{v}$ first computes its $L$ different LSH codes into the $L$ different hash table buckets. Then, a \textit{candidate set} of nearest neighbors of $\mathbf{v}$ is composed of the union $A(\mathbf{v})$ of all points hashed into the same $L$ buckets as $\mathbf{v}$. A \textit{short-list search} over $A(\mathbf{v})$ calculates the subset $k \subset A(\mathbf{v})$ of $k$ neighbor points that are closest in distance to $\mathbf{v}$. This short-list is typically implemented with a max-heap data structure and requires the most computation with LSH~\cite{Indyk:1998,Datar:2004,Pan:2010}.

Two surveys led by J. Wang et al.~\cite{Wang:2014,Wang:2016} review additional CPU-based techniques related to LSH. The following studies exclusively focus on data-parallel, GPU-based LSH approaches.
%provide a survey of hashing for similarity search which covers extant hashing algorithm research for LSH. Wang et al.~\cite{Wang:2016} survey an emerging area of \textit{learning to hash} that is based on LSH techniques and uses machine learning to estimate parameters of highly-accurate LSH hash functions; however, the majority of these techniques are beyond the scope of our survey. \fix{(Sam: I think you only need to say ``more CPU-based techniques are available from these two survey papers.'')} While most of the papers covered in these two surveys are oriented toward CPU usage, the following studies exclusively focus on data-parallel, GPU-based LSH hashing approaches.

Pan et al.~\cite{Pan:2010} design a data-parallel GPU-based variant of the canonical LSH method that simulates the $L$ different hash tables with a single cuckoo hash table of Alcantara et al.~\cite{Alcantara:2009}. In SIMT parallel fashion, threads execute insert and query operations on their assigned $D$-dimensional points. During an insertion, all of the LSH codes of points are data-parallel radix sorted in a linear array. Then, the sorted codes are partitioned into \textit{buckets}, based on unique code values. A data-parallel prefix-sum scan identifies the starting and ending indices of each bucket. The LSH code and its bucket interval define a key-value pair that is inserted into the cuckoo hash table, or \textit{indexing table}. Multiple points may have the same LSH code/key and thus collide at the same index table location. These collisions are resolved via a set of $L \geq 3$ hash functions that define the probe sequence needed to relocate points upon eviction. Note that the $L$ functions simulate hashing a point into the $L$ separate hash tables (or buckets) of traditional LSH. Finally, a query operation on a point computes the LSH code of the point, probes for the corresponding key in the indexing table, and then extracts the associated bucket interval value. The points within this bucket define the candidate set of neighbor points. A data-parallel search over a max-heap of these points returns the $k$ANN. Experiments on real-time motion planning data reveal that this approach is both faster and more accurate than comparable tree-based $k$ANN approaches. Also, the accuracy, or spatial locality, of the nearest neighbor hashing increases as the number of cuckoo hash functions $L$ increases.   

Pan and Manocha~\cite{Pan:2011} follow-up their original GPU-based LSH approach~\cite{Pan:2010} with \textit{bi-level} LSH that adds the following four components:
\begin{tightEnumerate}
\item \textit{Random projection (RP) tree}: In the first level, an RP tree is constructed in parallel over the input data points by iteratively partitioning the points into smaller and smaller clusters until a desired tree depth is reached, with leaf nodes containing small subsets of likely spatially-similar points. The tree is similar to a k-D tree, but splits the points along random directions instead of along coordinate directions~\cite{Dasgupta:2008}. This addition to the LSH helps generate more compact and accurate LSH codes.
\item \textit{Hiearchical LSH table}: In the second level, an LSH table is constructed for each different RP tree leaf node and its subset of points. Unlike the previous LSH approach, two additional steps are performed prior to computing LSH codes. First, each point in the leaf node is projected into a more compact $E_8$ lattice space that produces more accurate projections for high-dimensional data. Then, these LSH projections are mapped to 1D \textit{Morton curve} values that preserve neighborhood relationships. These values are efficiently constructed on the GPU, via the approach of Lauterbach et al.~\cite{Lauterbach:2009}, and serve as LSH codes.
\item \textit{Bi-level LSH code}: A modified Bi-level LSH code for a point $\mathbf{v}$ is specified by the pair $\widetilde{H} = ($RP-tree$(\mathbf{v}),H(\mathbf{v}))$, where RP-tree$(\mathbf{v})$ is the index of the leaf node to which $\mathbf{v}$ belongs and $H(\mathbf{v})$ is the Morton curve value (or LSH code). These bi-level codes are then data-parallel radix sorted and bucketed, producing the bucket intervals. Similar to the previous approach, the LSH codes $H(\mathbf{v})$ and their corresponding bucket intervals form key-value pairs for cuckoo indexing table.
\item \textit{Work queue}: Instead of extracting the $k$ANN of a query point $\mathbf{v}$ from a global memory max-heap of size $k$, a global memory work-queue is used to perform a clustered sort that arranges the candidate set of neighbors in ascending order of distance from $\mathbf{v}$. This sort works in data-parallel across multiple queries and candidate sets, and employs radix sorting, which can benefit from the high-speed GPU shared memory. Moreover, this queueing approach increases parallel throughput and avoids thread branch divergence inherent in the max-heap tree traversals.
\end{tightEnumerate}
A set of experiments on an image dataset with nearly 2 million images, each represented as a $512$-dimensional point, demonstrate that the Bi-level LSH can provide higher quality ANN results than the previous LSH method, given the same computational budget. Each of the algorithmic improvements discussed above attain accelerated GPU performance over the original methods.   

Gieseke et al.~\cite{Gieseke:2014} introduce a \textit{buffer k-d tree} data structure for massively-parallel ANN search on the GPU. While hashing is not used in this study, the authors state a weakness of the Pan and Manocha~\cite{Pan:2011} approach is that it possibly yields inexact answers, as compared to those of a serial benchmark. While a reason was not provided, this inaccuracy may be due to either the RP-tree spatial partitioning of points or the hiearchical LSH code calculation, which involves consecutive mappings to lower-dimensional spaces. 

Luka{\v{c}} and {\v{Z}}alik~\cite{Lukac:2015} implement a GPU-optimized variant of \textit{multi-probe LSH} (MLSH) that was originally introduced by Lv et al.~\cite{Lv:2007}. In this approach, $L$ hash tables are constructed, one at a time, on the GPU using the unique LSH codes of projected multi-dimensional points as bucket indices (the LSH code is a function of $K$ different LSH projections to a line). The points within each bucket are sorted in ascending order by ID using the data-parallel radix sort of Merrill et al.~\cite{Merrill:2010}. During point queries, the candidate sets are composed using \textit{query-directed probing} to first visit buckets that possess a high success probability of containing nearest neighbors. Given the properties of LSH, these neighbors are likely to be in buckets that only differ slightly in distance in the table. A \textit{scoring criteria} assigns a threshold for each bucket, determining whether it should be probed, based on its likelihood of containing a nearest neighbor of the query point. In this work, a simplified multi-probe scheme assigns a scoring criteria to the immediate left and right buckets of the bucket into which the query point is hashed. Thus, the points of at most 3 buckets combine to form the candidate set. In order to quickly extract the $k$ANN for the query point, a \textit{deterministic skip-list} (DSL) search structure is built in global memory. This structure arranges the candidate set points in multiple levels of sorted linked lists, or subsequences, each of increasing size and in order of distance from the query point~\cite{Munro:1992,Moscovici:2017}. The resulting $k$ANN is copied back to host CPU memory, and then the LSH procedure is iteratively repeated for the remaining $L-1$ hash tables, after which $L$ different sets of $k$ANN reside on the host. From these $L\cdot k$ points, a final $k$ANN is determined. 

\subsection{Separate Chaining} 
\label{separate-chaining}

\textit{Separate chaining} is a classic collision resolution technique that uses a linked list or node-based data structure to store multiple collided keys at a single hash table entry. Each hash table entry contains a pointer, or memory address, to a head node of a linked list, or \textit{chain}. Each node in the linked list consists of a key, associated value (optional), and a pointer to the next node in the list, if any. If a single key hashes to an entry, then the linked list consists of a single node with a null pointer to the non-existent next node. Otherwise, if multiple keys collide and hash to the same location, then the linked list forms a \textit{chain} of these keys, each represented by a separate node in the list. During a query operation, a key hashes to an entry in the table and then iterates through each of the nodes of the chain referenced at the entry, searching for a matching key. This iterative search is similar in nature to open-addressing linear probing (refer to subsection~\ref{linear-probing}), where a key hashes to an initial table entry and then probes each subsequent entry until a matching key is found. Both techniques can result in degenerate, worst-case queries that require a non-constant number of probes. Unlike separate chaining, linear probing is prone to primary clustering of collided keys and performs \textit{lazy deletion} of keys that renders unoccupied table entries heavily fragmented and may require re-hashing or compaction. However, linear probing is more amenable to caching, as probes are conducted within a contiguous block of memory, instead of over the scattered memory of linked list nodes.     

Moreover, separate chaining is a form of \textit{open hashing} in which keys and values are stored in allocated linked lists outside of the hash table and then referenced by head node pointers that are stored inside the table. Contrarily, open addressing collision resolution follows \textit{closed hashing}, whereby each hashed key (and value) is inserted directly into the hash table.

In the context of parallel hashing, separate chaining must synchronize collisions during key insertions to ensure that the linked list data structures are properly allocated and constructed. Moreover, a dynamic memory allocation scheme must ensure that concurrent threads conducting insert operations properly synchronize their requests for new available blocks of memory. Similar design challenges exist for the deletion of keys, and the simultaneous execution of queries by threads must avoid reader-writer race conditions that result in faulty memory accesses to incorrect or deallocated nodes (keys).

A large body of research has investigated \textit{concurrent} hash tables for separate chaining on multi- and many-core CPU systems~\cite{Greenwald:2002,Michael:2002,Peierls:2005,Shalev:2006,Goodman:2010}. Each of these hash tables is designed to support dynamic\footnotemark updates and resizing with lock-based methods (e.g., mutexes or spin-locks) or lock-free (non-blocking) hardware atomics, such as compare-and-swap (CAS). Since the majority of these hash tables are linked list-based data structures, they are not designed for scalable, high-performance on massively-parallel GPU architectures. In particular, when ported to the GPU, the performance of these approaches may degrade due to several reasons:
\begin{tightItemize}
\item Lock-based methods induce substantial thread contention during blocking operations for shared resources and are not scalable with increasing numbers of concurrent threads. This contention creates starvation for blocked threads and warp underutilization, since each thread must wait for its other warp threads to finish acquiring and releasing the lock. Moreover, lock-free hardware atomic primitives prevent deadlock, but still neglect the sensitivity of GPUs to global memory accesses and thread branch divergence.
\item Lack of coalescing among memory accesses due to the scattering of linked list node pointers in memory and random addressing of keys by threads within the same warp, which can lead to additional global memory transactions (cache line loads).
\item Dynamic memory management and pointer chasing required for the linked lists on the GPU is challenging for traditional CPU-based synchronization schemes, due to the massive thread parallelism. This performance challenge is similarly observed in pointer-heavy spatial search tree structures that are ported to the GPU. 
\end{tightItemize}
The following studies introduce GPU-based separate chaining hashing approaches that attempt to address these performance challenges.  

\footnotetext{Some implementations are aware of future insertions at compile-time and preallocate sufficiently-large additional memory. These hash tables are \textit{semi-dynamic} since they do not dynamically allocate new memory at runtime for unknown insertions.}

Moazeni and Sarrafzadeh~\cite{Moazeni:2012} and Misra and Chaudhuri~\cite{Misra:2012} deploy some of the earliest lock-free, separate chaining-based hash tables on a GPU architecture. Using CUDA atomic CAS operations (atomicCAS and atomicInc), both approaches support batches of concurrent query and insert operations, with only~\cite{Misra:2012} also supporting delete operations. \cite{Moazeni:2012} achieves a significant execution time speedup for queries over counterpart lock-based and OpenMP-based CPU implementations. However, the lock-free table only attains significantly higher throughput (operations per second) than the OpenMP implementation for query-heavy batches ($80\%$ queries and $20\%$ inserts). Additionally, this work does not focus on larger, scalable batch sizes and provides little analysis regarding thread- or warp-level performance. \cite{Misra:2012} demonstrates that a GPU lock-free hash table leverages a much higher degree of concurrency and throughput than a CPU implementation for both query-heavy and update-heavy workload batches. This performance increase is largely due to spreading the thread contention and atomic comparisons over multiple different hash locations, as threads work in SIMT data-parallel fashion to conduct mixed operations at random locations.

Stuart and Owens~\cite{Stuart:2011} and newer versions of the Nvidia CUDA C library~\cite{Nvidia-ProgramGuide} both introduce new efficient synchronization and atomic primitives (e.g., warp-level and share memory atomics) for CUDA-compatible GPUs. These primitives likely satisfy the inefficiencies of atomics for pointer-based data structures cited in Misra and Chaudhuri~\cite{Misra:2012}.

Steinberger et al.~\cite{Steinberger:2012} design \textit{ScatterAlloc}, an efficient GPU-based dynamic memory allocator that is significantly faster than the built-in CUDA toolkit allocator and the first-published GPU allocator, \textit{XMalloc}, of Huang et al.~\cite{Huang:2010}. This scheme maintains a linked-list memory pool of \textit{super blocks} and organizes the blocks into larger fixed-size \textit{pages}, which are addressed via a hash function. For usage in separate chaining hashing on the GPU, linked list collision chains can be dynamically allocated or deallocated as super blocks to large numbers of threads in parallel, as part of update operations (e.g., insert or delete). Due to the hash-based addressing of available memory pages, threads can minimize contention for the same block of memory and scatter their block assignments for efficient random access (with a possible tradeoff of memory fragmentation). Vinkler and Havran~\cite{Vinkler:2015} survey and experimentally compare existing GPU dynamic memory allocation schemes. The performance of each scheme varies across different criteria, including fragmentation of available memory blocks, per-block thread contention for atomic allocation requests, size and coalescing of requested memory by inter-warp threads, uniformity of the number of allocation requests per inter-warp thread, and dependence on the number of user-specified registers available to threads in each SM of the GPU.

Ashkiani et al.~\cite{Ashkiani:2017} propose a dynamic \textit{slab hash} table on the GPU that is built upon an array of linked-lists, or \textit{slab lists}, each of which represent a chain of one or more \textit{slabs}, or memory units, that store collided keys. Each slab of memory is roughly the size of a warp memory transaction width (128 bytes), or the number of warp threads (32) times the size of a key (4 bytes). Thus, each warp is aligned to perform operations over the keys stored in a single slab. As part of a novel \textit{work-cooperative work sharing} (WCWS) strategy, each warp maintains a \textit{work queue} that stores all the arbitrary operations assigned to the different threads in the warp. In a round-robin fashion, each batch of the same operation type in the queue is fully and cooperatively executed by the threads. For a given operation type, all threads perform a warp-wide ballot instruction to denote the \textit{active} threads that were assigned this operation. For each active thread, the entire warp cooperates to execute the active thread's operation and its corresponding key. If the operation is a query for a key $q$, then the warp hashes $q$ to a slab list $b_i = h(q)$ in the slab hash table $H$. The first warp-sized slab, $b_{i0}$, of the slab list at $H[b_i]$ is loaded from global memory via a single memory transaction. This slab memory unit contains the same number of keys as threads in the warp. So, each warp thread then compares its corresponding key $k$ with the query key $q$. If any thread has $k = q$, then a successful result is returned. Otherwise, the warp follows a \textit{next} pointer stored in $b_{i0}$ to load the next connected slab $b_{i1}$, in which $q$ is cooperatively searched again. This search ends when either $q$ is found or the last slab in $b_i$ has been searched. 

An insert operation proceeds similarly, except the threads search for an empty slab spot into which a new key can be atomically inserted. If no empty spot is found in any of the slabs of the slab list, then a new slab must be atomically and dynamically allocated (since other warps may also be trying to allocate). This allocation is efficiently performed via a novel warp-synchronous \textit{SlabAlloc} allocator (see~\cite{Ashkiani:2017} for further details).  

This warp-cooperative approach differs from previous GPU separate chaining in which the threads of a warp execute a SIMT query or update operation for different keys, each of which likely require a random, uncoalesced memory access. WCWS ensures memory coalescing for each operation by perfectly aligning the threads of a warp with the keys of a slab, both of the same size. Thus, the same block of cache-aligned global memory is loaded in a single transaction for every operation by the warp, exposing increased throughput (millions of operations per second). Moreover, while being inserted, keys are always stored at contiguous addresses within a slab memory unit. This contrasts with traditional linked list storage in which keys are inserted as new nodes at random memory locations.

The performance of the dynamic slab hash table is compared to the static cuckoo hash table of Alcantara et al.~\cite{Alcantara:2009}---which must be rebuilt upon updates---and the semi-dynamic lock-free hash table of Misra and Chaudhuri~\cite{Misra:2012}. For static bulk builds, cuckoo hashing consistently achieves a higher throughput of insertions per second, while slab hashing achieves higher query throughput only when the average number of slabs per slab list is less than 1 (i.e., approximately a single ``node'' list). Over all configurations, cuckoo hashing attains the better query throughput. In the best case scenario, it only makes a single atomic comparison for an insertion and a single random memory access for a query; contrarily, in the best case, slab hashing requires both a memory access (to load a slab) and an atomic comparison for an insertion. For dynamic updates, slab hashing significantly outperforms cuckoo hashing, in terms of execution time, as the number of inserted keys increases. This is due to the rebuilding of the static cuckoo hash table each time a new batch is inserted. Additionally, slab hashing significantly outperforms lock-free hashing across different distributions of mixture operations and increasing numbers of slab lists (i.e., the size of the hash table).   

\section{Hashing Applications}
\label{applications}
The following section highlights a variety of real-world applications of the GPU-based hashing techniques presented in this survey. These applications can be broadly divided into six categories, many falling within the domains of computer graphics and database processing. The majority of the studies cited within each application area also introduce a novel hashing technique and are discussed in section~\ref{hashing-techniques}; the remaining studies strictly apply one of the hashing techniques. 

\textit{Collision detection}: Teschner et al.\cite{Teschner:2003} and Eitz and Lixu~\cite{Eitz:2007} use spatial hashing to detect real-time intersections between deformable objects in a scene and tetradedral cells in 3D mesh volumes. Lefebvre and Hoppe~\cite{Lefebvre:2006} use perfect spatial hashing to detect collisions among surfaces of 3D objects. Pouchol et al.~\cite{Pouchol:2009} use spatial hashing to model the interaction between solid objects (e.g., spheres) and fluid. Choi et al.~\cite{Choi:2009} use perfect spatial hashing to detect interference between characters and obstacles in a free space mapped virtual environment. Chentanez et al.~\cite{Chentanez:2016} use spatial hashing to detect and delete overlapping, or intersecting, triangles on the surface of 3D mesh volumes.

\textit{Adaptive mesh refinement} (AMR): Tumblin et al.~\cite{Tumblin:2015} use compact perfect hashing to search for neighboring cells in cell-based AMR for a shallow-water hydrodynamics simulation (e.g., AMR at the boundary of a water wave). Chentanez et al.~\cite{Chentanez:2016} use spatial hashing to perform AMR on 3D mesh volumes, as vertices are advected in real-time.

\textit{Surface rendering}: Lefebvre and Hoppe~\cite{Lefebvre:2006} use perfect spatial hashing to render the color surfaces of 3D volumetric textures. Alcantara et al.~\cite{Alcantara:2009,Alcantara:2011} (open-addressing cuckoo hashing), Garcia et al.~\cite{Garcia:2011} (open-addressing robin hood hashing), Nie{\ss}ner et al.~\cite{Niener:2013} (spatial hashing), and Duan et al.~\cite{Duan:2017} (spatial hashing) all perform real-time surface rendering and reconstruction of 3D volumes within voxelized grids. Bastos and Celes~\cite{Bastos:2008} use perfect hashing to perform isosurface rendering and morphing of adaptively sampled distance fields (ADFs). K{\"a}hler et al.~\cite{Kahler:2016} use spatial hashing to render voxelized 3D scene models of signed distance fields (SDFs).  

\textit{Interactive drawing and painting}: Lefebvre and Hoppe~\cite{Lefebvre:2006} use perfect spatial hashing to interactively paint over 3D volumetric textures. Garcia et al.~\cite{Garcia:2011} use open-addressing robin hood hashing to interactively draw on 2D surfaces, such as an atlas. Eyiyurekli and Breen~\cite{Eyiyurekli:2011} use spatial hashing to interactively edit and draw over 3D level-set surfaces.

%\textit{Duplicate detection}: \cite{Lessley:EGPGV2016} and~\cite{Lessley:LDAV17-1} use hash fighting to detect and remove duplicate internal faces of tetrahedral cells in 3D mesh volumes, leaving only the external, surface faces.

\textit{Database processing}: Hetherington et al.~\cite{Hetherington:2015} and Choudhury et al.~\cite{Choudhury:2016} use open-addressing cuckoo hashing to cache most-recently used, or working set, queries in a key-value store. Karnagel et al.~\cite{Karnagel:2015} use open-addressing linear probing to perform group-by and aggregation queries from a key-value store. Zhang et al.~\cite{Zhang:2015} and Breslow et al.~\cite{Breslow:2016} use open addressing bucketized cuckoo hashing to accelerate queries and updates in key-value stores. 

\textit{Similarity search}: Zhou et al.~\cite{Zhou:2016} use open-addressing robin hood hashing to extract the top-$k$ most similar matches for query records in real-world document and relational datasets. Alcantara et al.~\cite{Alcantara:2009} use open-addressing cuckoo hashing to perform geometric hashing, which is a form of 2D image matching. Pan et al.~\cite{Pan:2010}, Pan and Manocha~\cite{Pan:2011}, and Luka{\v{c}} and {\v{Z}}alik~\cite{Lukac:2015} each use locality-sensitive hashing to find the $k$ approximate nearest neighbors ($k$ANN) of query points within multi-dimensional record sets. Pouchol et al.~\cite{Pouchol:2009} use spatial hashing to perform particle neighbor search within fluid and solid interaction simulations. Todd et al.~\cite{Todd:2016} use multi-level bucketized hashing to identify genes with similar $k$-motifs, or DNA subsequences of length $k$.
\section{Analysis and Future Work}
\label{analysis}

\begin{table*}[ht!]
   \centering
    \begin{tabular}{|l||p{2.2cm}|p{1.5cm}|p{2cm}|p{1.3cm}|}
    \hline
    Use Case Attribute/Hashing Category &  Open-Addressing & Perfect Hashing & Spatial Hashing & Separate Chaining \\ \hline \hline \hline
    \textbf{\textit{Access Patterns}:} & ~               & ~               & ~               & ~  \\ \hline
    -- Ordered queries  & \textit{CoherentHash}\cite{Garcia:2011} & ~   & ~               & ~  \\
    -- Random insertions and queries                       & \textit{CuckooHash2}\cite{Alcantara:2011}               & ~               & ~               & ~  \\
    -- Duplicate insertions and queries                       & ~               & \textit{PerfectHash}\cite{Lefebvre:2006}               & \textit{EGSH}\cite{Duan:2017}              &   \\
    -- Query-heavy operation mix & \textit{HortonHash}\cite{Breslow:2016}; \textit{MemcachedGPU}\cite{Hetherington:2015} & & & \\
    
    -- Update-heavy operation mix & & & & \textit{SlabHash} \cite{Ashkiani:2017} \\
    
    -- Unsuccessful (empty) queries & & \textit{PerfectHash}\cite{Lefebvre:2006} & & \\
    \hline \hline
    \textbf{\textit{Data Type}:}                       & ~               & ~               & ~               & ~  \\ \hline
    -- Grid-based spatial primitives                       & ~               & ~               & \textit{VoxelHash}\cite{Niener:2013}              & ~ \\
    -- Integer or index-based                       &  \textit{CuckooHash2}\cite{Alcantara:2011}              & ~               & ~               & ~ \\
    -- Multi-dimensional attribute vector                       & ~               & ~               & \textit{BiLevelLSH}\cite{Pan:2011}               &  \\ \hline \hline
    \textbf{\textit{Hash Table}:}                       & ~               & ~               & ~               & ~  \\ \hline
    -- Collision-free & & \textit{PerfectHash}\cite{Lefebvre:2006} & & \\
    -- Fast construction & \textit{CuckooHash2}\cite{Alcantara:2011} & & & \\
    
    -- Dynamic & & & & \textit{SlabHash} \cite{Ashkiani:2017} \\
    
    -- Low occupancy                       & ~               & ~               & \textit{CompactHash}\cite{Tumblin:2015}               & ~  \\
    -- High occupancy; maximum load                       & \textit{CoherentHash}\cite{Garcia:2011}              & ~               & ~               & ~  \\ \hline \hline
    \textbf{\textit{Hardware:}}                       & ~               & ~               & ~               & ~  \\ \hline
    -- Use of CPU memory (PCIe bound)                       & \textit{StadiumHash}\cite{Khorasani:2015}; \textit{HortonHash}\cite{Breslow:2016}               & ~               & ~               & ~    \\
    -- Use of GPU shared memory & \textit{CuckooHash1}\cite{Alcantara:2009} & & \textit{BiLevelLSH}\cite{Pan:2011} & \\
    -- Efficient use of atomics & \textit{CuckooHash2}\cite{Alcantara:2011} & & & \\ \hline
    \end{tabular}
    \caption {Suggested hashing technique(s) for different use case attributes. For each attribute, the most suitable or best-performing technique from one or more of the four hashing categories is denoted. Additional details regarding a technique can be found within the section of its encompassing hashing category.}
    \label{table:use-cases}
\end{table*}

\begin{table*}[t!]
   \centering
    \begin{tabular}{|l||>{\centering\arraybackslash}p{1.5cm}|>{\centering\arraybackslash}p{1.5cm}|>{\centering\arraybackslash}p{1cm}|>{\centering\arraybackslash}p{1.9cm}||>{\centering\arraybackslash}p{1.1cm}|>{\centering\arraybackslash}p{1.4cm}|>{\centering\arraybackslash}p{1.4cm}|}
    \hline
    & \multicolumn{4}{ c|| }{GPU Performance Criteria} & \multicolumn{3}{ c| }{GPU Hardware Features} \\ \hline \hline
    Hashing Technique & Sufficient Parallelism & Memory Coalescing & Control Flow & CPU$\leftrightarrow$GPU Data Transfers & Shared Memory & Atomic Operations & Warp-wide Voting \\ \hline \hline \hline
    \textbf{\textit{Open-addressing}:} & & & & & & &\\ \hline
    -- \textit{CoherentHash}\cite{Garcia:2011} & \ding{51} & \ding{51} & \ding{53} & \ding{53} & \ding{53} & \ding{51} & \ding{53}\\
    -- \textit{CuckooHash1}\cite{Alcantara:2009} & \ding{51} & \ding{53} & \ding{53} & \ding{53} & \ding{51} & \ding{51} & \ding{53}\\
    -- \textit{CuckooHash2}\cite{Alcantara:2011} & \ding{51}  & \ding{53} & \ding{53} & \ding{53} & \ding{53} & \ding{51} & \ding{53}\\
    -- \textit{HortonHash}\cite{Breslow:2016} & \ding{51}  & \ding{53} & \ding{51} & \ding{51} & \ding{53} & \ding{53} & \ding{53}\\
    -- \textit{MemcachedGPU}\cite{Hetherington:2015} & \ding{51}  & \ding{53} & \ding{53} & \ding{51} & \ding{51} & \ding{51} & \ding{53}\\
    -- \textit{StadiumHash}\cite{Khorasani:2015} & \ding{51}  & \ding{53} & \ding{53} & \ding{51} & \ding{51} & \ding{51} & \ding{51}\\
    \hline \hline
    \textbf{\textit{Perfect Hashing}:} & & &  &  & & & \\ \hline
    -- \textit{PerfectHash}\cite{Lefebvre:2006} & \ding{51} & \ding{51} & \ding{51} & \ding{51} & \ding{53} & \ding{53} & \ding{53}\\ \hline \hline
    \textbf{\textit{Spatial Hashing}:}  &  & & & & & &\\ \hline
    -- \textit{BiLevelLSH}\cite{Pan:2011} & \ding{51} & \ding{53} & \ding{53} & \ding{53} & \ding{51} & \ding{51}& \ding{53}\\
    -- \textit{EGSH}\cite{Duan:2017} & \ding{51} & \ding{53} & \ding{51} & \ding{51}& \ding{53} & \ding{53} & \ding{53}\\
    -- \textit{VoxelHash}\cite{Niener:2013} & \ding{51} & \ding{53} & \ding{53} & \ding{53} & \ding{53} & \ding{51} & \ding{53}\\ \hline \hline
    \textbf{\textit{Separate Chaining}:} & & & & & & &\\ \hline
    -- \textit{SlabHash} \cite{Ashkiani:2017} & \ding{51} & \ding{51} & \ding{51} & \ding{53} & \ding{53} & \ding{51} & \ding{51}\\ \hline
    \end{tabular}
    \caption {
    Select hashing techniques and their ability to address GPU criteria for optimal performance and utilize performant GPU hardware features. The techniques are grouped by category and represent the subset of techniques that are identified as highly-suitable for different use-case attributes in Table~\ref{table:use-cases}.
    }
    \label{table:performance-criteria}
\end{table*}

This section analyzes the findings of the surveyed hashing techniques and identifies opportunities for future work. Table~\ref{table:use-cases} enumerates a set of $17$ hashing use case attributes and suggests the most-suitable or performant hashing technique(s) for each attribute. Due to the large number of possible subsets of use case attributes, a technique is only suggested for a single attribute. A practitioner can consult the table for a set of desired attributes, identify overlapping suggested techniques, and then investigate the suitability of these techniques for a specific task. Table~\ref{table:performance-criteria} evaluates the most-suitable hashing techniques from Table~\ref{table:use-cases} based on their ability to address optimal GPU performance criteria and utilize performant GPU hardware features. This evaluation assesses performance as it pertains to arbitrary access patterns for insertions and queries. Thus, special cases such as empty queries or ordered accesses are not considered unless a technique is specifically designed to perform well for such cases; for example, \textit{CoherentHash}~\cite{Garcia:2011} achieves best-in-class throughput and memory coalescing among open-addressing techniques only when coherence exists among input elements and their hash table locations. The GPU performance criteria and hardware features are described as follows:
\begin{tightItemize}
\item Sufficient Parallelism: The hashing technique experimentally demonstrates a sufficient throughput of insertion and query operations (millions per second) to hide global memory access latency.   
\item Memory Coalescing: All the threads in a warp access addresses within the same fetched cache line of contiguous memory. These memory requests are necessary to execute the given SIMT instruction.
\item Control Flow: All the threads in a warp follow the same execution path for a SIMT instruction.
\item CPU$\leftrightarrow$GPU Data Transfers: The hash table is constructed and/or stored in CPU memory and then accessed from or copied onto the GPU via the interconnection bus (e.g., PCI-e); thus, the hashing experiences data transfer bandwidth latency.
\item Shared Memory: Per-thread-block GPU memory space that is smaller in size than global DRAM memory, but offers faster memory accesses.
\item Atomic Operations: Lightweight hardware atomic functions, such as compare-and-swap (CAS), that guard and manage hash table entries during parallel insertions, probing evictions (e.g., in cuckoo hashing), and deletions.
\item Warp-wide Voting: Lightweight functions used by all the threads in a warp to communicate data and perform collaborative execution, such as when all warp threads query the hash table for the same key.  
\end{tightItemize}

For arbitrary, random access patterns, \textit{CuckooHash2} cuckoo hashing~\cite{Alcantara:2011} offers best-in-class throughput performance among the surveyed hashing techniques (subsection~\ref{cuckoo-hashing}). This is due to the small constant number of probes necessary in both the best- and worst-case scenarios. In the worst-case insertion scenario of not finding an empty slot, the cuckoo hash table demonstrates fast reconstruction rates. In the presence of spatially-ordered access patterns, the \textit{CoherentHash} robin hood hashing~\cite{Garcia:2011} achieves greater throughput than cuckoo hashing and is robust to higher load factors (subsection~\ref{robin-hood-hashing}). 

In the ideal, ``fast-path,'' scenario, an open-addressing technique only requires a single atomic CAS operation for an insertion and a single random global memory access for a query. However, in a typical scenario, a variable number of probes are needed to insert and query a key, often spanning non-contiguous regions of memory. This induces non-coalesced memory accesses and control flow divergence among threads of a warp. Thus, most of the open-addressing techniques assessed in Table~\ref{table:performance-criteria} cannot guarantee to attain memory coalescing and control flow.

The combination of radix sorting and binary searching is a very effective alternative to searching via hashing when access patterns are ordered or the data is already in near-sorted order prior to sorting. However, for interactive use, this approach naively requires a re-sort of a larger array each time new data is added. Additional research is needed to investigate more-efficient data-parallel schemes for accommodating dynamic data.

If data will be updated at run-time, then \textit{SlabHash}~\cite{Ashkiani:2017} offers best-in-class dynamic hashing, achieving a significant increase in throughput over cuckoo hashing, which must be reconstructed after each batch of updates (section~\ref{separate-chaining}). Moreover, as seen in Table~\ref{table:performance-criteria}, this technique addresses each of the criteria for optimal GPU performance. Further research is needed to compare the performance of slab hashing with that of \textit{CoherentHash} robin hood hashing~\cite{Garcia:2011} in the presence of coherent access patterns.   

When data must be stored and accessed off-device in CPU memory, the use of ticketing, or key bit signatures, is beneficial to save expensive accesses for obvious non-matches during probing/querying. Future hashing approaches should assess the performance benefits of ticketing even when off-device accesses do not occur. Maintaining the ticketing structure in shared memory appears to be particularly beneficial, as demonstrated by the \textit{StadiumHash} open-addressing technique~\cite{Khorasani:2015}.

Regardless of the data use case, shared memory should be leveraged as much as possible to perform warp operations and faster memory accesses (not necessarily coalesced). This is facilitated by sizing buckets to the size of a thread block, such as in \textit{CuckooHash1} cuckoo hashing~\cite{Alcantara:2009}. If data must be accessed outside of shared memory, warps should be modeled as collaborative processing units the size of a memory transaction. Each thread is assigned to an entry within the loaded cache line and all threads then compare their entries (possibly empty) to the query or insert key via a warp-wide voting function. \textit{CuckooHash1}~\cite{Alcantara:2009}, \textit{StadiumHash}~\cite{Khorasani:2015}, and \textit{SlabHash}~\cite{Ashkiani:2017} make particularly good use of shared memory and warp-wide voting (Table~\ref{table:performance-criteria}).  

Fast hash table construction enables larger load factors, acceptance of insertion failure, and dynamic usage in interactive applications. CPU-constructed hash tables face two bottlenecks: slower construction on the CPU and copying over the PCIe bus to the GPU. Both bottlenecks render these tables infeasible for updates or reconstructions. From Table~\ref{table:performance-criteria}, the \textit{HortonHash}~\cite{Breslow:2016}, \textit{PerfectHash}~\cite{Lefebvre:2006}, and \textit{EGSH}~\cite{Duan:2017} techniques are bandwidth-bound by the transfer of the hash table from CPU to GPU prior to querying. Additionally, \textit{MemcachedGPU}~\cite{Hetherington:2015} and \textit{StadiumHash}~\cite{Khorasani:2015} must service data transfers during querying, as hash table data resides on the CPU. Further research is needed to redesign CPU-constructed hash tables for efficient data-parallel construction on the GPU.    

Perfect hashing (section~\ref{perfect-hashing}), \textit{PerfectHash}~\cite{Lefebvre:2006}, avoids collision resolution, but is not well-suited for updates, since the hash table must be reconstructed on the CPU and remain PCIe bandwidth-bound. A trade-off arises: either use multiple separate hash tables (and multiple probes), or use a single addressable hash table and construct the offset table, which is the primary bottleneck during construction. Further research towards constructing the offset table in data-parallel on the GPU is needed to make perfect hashing a more dynamic, interactive solution.

Compact spatial hashing (subsection~\ref{sec:grid-based}), \textit{CompactHash}~\cite{Tumblin:2015}, offers the useful feature of downsizing a perfect hash table that contains a significant number of unused entries, which arises often in spatial hashing. This comes with the trade-off of new hash collisions that must be resolved. Further research should assess the viability of this approach for other types of hash tables and varying load factors.

The \textit{BiLevelLSH}~\cite{Pan:2011} locality-sensitive hashing technique takes advantage of fast on-device data-parallel operations to sort key-value pairs and hash them into a cuckoo hash table. Further work is needed to design a dynamic variant that supports updates to the hash table and sorted key-values. Moreover, future research should investigate the use of LSH for \textit{approximate} surface rendering and reconstruction tasks. For instance, instead of querying the data to render for each point in a grid, select points can be queried and return, in a single operation, the approximate data for an entire $k$-point bounding box in the form of the $k$ANN. 

Finally, prospective avenues for future research exist for a \textit{HashFight} technique that is introduced by Lessley et al.~\cite{Lessley:EGPGV2016,Lessley:LDAV17-1} as part of a platform-portable, GPU-compatible hashing approach. This approach employs an iterative winner-takes-all collision resolution technique that does not use hardware atomic primitives to synchronize writes to the hash table. Instead, race conditions are a fundamental and non-detrimental feature of resolving collisions. However, \textit{HashFight} does not maintain a persistent hash table, but rather reconstructs a new, smaller-sized table at the beginning of each iteration. Thus, additional work is needed to expand the technique to support query and insert operations, with accompanying throughput analyses. Then, the build speed of \textit{HashFight} can be compared against the build times of the best-in-class static cuckoo and robin hood hashing techniques, particularly \textit{CuckooHash2}~\cite{Alcantara:2011} and \textit{CoherentHash}~\cite{Garcia:2011}.   

%Hash fighting (section~\ref{hash-fighting}) is based on a non-persistent hash table that is repetitively and quickly reconstructed without the need for atomic primitives. This technique presents several prospective avenues for further research. First, the build speed should be compared against the best-in-class build times of the static cuckoo and robin hood hash tables. Second, additional work is needed to expand the technique to support query and insert operations, with accompanying throughput analyses. An initial direction towards supporting queries is to append them to the original set of build keys and then perform hash fighting until all threads have completed their queries; in the case of empty queries, a maximum threshold of iterations would need to be set before a negative result is returned. A query would return positive if it has a duplicate key in the set of build keys. Insertions could be supported in a similar appending fashion.

%Important findings from this survey and Table~\ref{table:use-cases} are summarized as follows, along with accompanying avenues of future research.

%Notable observations from Table~\ref{table:performance-criteria} are as follows. First, the \textit{SlabHash}~\cite{Ashkiani:LFAO17} technique addresses each of the criteria for optimal GPU performance  

\section{Conclusion}
\label{conclusion}

This paper provides a survey of parallel hashing techniques for GPU architectures. These techniques are categorized according to the method of collision resolution: open-addressing, perfect hashing, spatial hashing, and separate chaining. Each of the surveyed studies offer various design choices and patterns that help inform a more-general set of best practices for performant hashing on the GPU. These best practices and the most-suitable hashing techniques for different use-case factors are analyzed and used to reveal opportunities for future research. 

%%%%%%%%%%%%%%%%%%%%%%%%%%%%%%%%%%%%%%%%%%%%%%%%%%%%%%%%%%%%%%%%%%%%%%%%%%%%%%%%

\bibliographystyle{abbrv-doi-hyperref-narrow}

\bibliography{refs}

\begin{thebibliography}{100}
\renewcommand*{\sfdefault}{PTSansNarrow-TLF}

\bibitem{cuddp}
{CUDA Data Parallel Primitives Library}.
\newblock \url{http://cudpp.github.io}, Nov. 2017.

\bibitem{vtk-m}
{VTK-m}.
\newblock \url{https://gitlab.kitware.com/vtk/vtk-m}, Nov. 2017.

\bibitem{Alcantara:2009}
D.~A. Alcantara, A.~Sharf, F.~Abbasinejad, S.~Sengupta, M.~Mitzenmacher, J.~D.
  Owens, and N.~Amenta.
\newblock Real-time parallel hashing on the gpu.
\newblock In {\em ACM SIGGRAPH Asia 2009 Papers}, SIGGRAPH Asia '09, pp.
  154:1--154:9. ACM, New York, NY, USA, 2009.

\bibitem{Alcantara:2011}
D.~A. Alcantara, V.~Volkov, S.~Sengupta, M.~Mitzenmacher, J.~D. Owens, and
  N.~Amenta.
\newblock Chapter 4 - {B}uilding an efficient hash table on the \{GPU\}.
\newblock In W.-m.~W. Hwu, ed., {\em \{GPU\} Computing Gems Jade Edition},
  Applications of GPU Computing Series, pp. 39 -- 53. Morgan Kaufmann, Boston,
  2012.

\bibitem{Amdahl:1967}
G.~M. Amdahl.
\newblock Validity of the single processor approach to achieving large scale
  computing capabilities.
\newblock In {\em Proceedings of the April 18-20, 1967, Spring Joint Computer
  Conference}, AFIPS '67 (Spring), pp. 483--485. ACM, New York, NY, USA, 1967.

\bibitem{Ashkiani:2016}
S.~Ashkiani, A.~Davidson, U.~Meyer, and J.~D. Owens.
\newblock Gpu multisplit.
\newblock In {\em Proceedings of the 21st ACM SIGPLAN Symposium on Principles
  and Practice of Parallel Programming}, PPoPP '16, pp. 12:1--12:13, 2016.

\bibitem{Ashkiani:2017}
S.~{Ashkiani}, M.~{Farach-Colton}, and J.~D. {Owens}.
\newblock {A Dynamic Hash Table for the GPU}.
\newblock In {\em Proceedings of the 31st IEEE International Parallel and
  Distributed Processing Symposium}, IPDPS '18, pp. 419--429, May 2018.

\bibitem{Ashkiani:LFAO17}
S.~Ashkiani, S.~Li, M.~Farach{-}Colton, N.~Amenta, and J.~D. Owens.
\newblock {GPU} {LSM:} {A} dynamic dictionary data structure for the {GPU}.
\newblock In {\em Proceedings of the 31st IEEE International Parallel and
  Distributed Processing Symposium}, IPDPS '18, pp. 430--440, May 2018.

\bibitem{Bastos:2008}
T.~Bastos and W.~Celes.
\newblock Gpu-accelerated adaptively sampled distance fields.
\newblock In {\em 2008 IEEE International Conference on Shape Modeling and
  Applications}, pp. 171--178, June 2008.

\bibitem{blelloch1990vector}
G.~E. Blelloch.
\newblock {\em Vector models for data-parallel computing}, vol.~75.
\newblock MIT press Cambridge, 1990.

\bibitem{Blikberg:2005}
R.~Blikberg and T.~Sørevik.
\newblock Load balancing and openmp implementation of nested parallelism.
\newblock {\em Parallel Computing}, 31(10):984 -- 998, 2005.
\newblock OpenMP.

\bibitem{BoostIterator:web:2003}
{Boost C++ Libraries}.
\newblock {Boost.Iterator Library}, 2003.
\newblock
  \url{http://www.boost.org/doc/libs/1_65_1/libs/iterator/doc/index.html}.

\bibitem{Bordawekar:2014}
R.~Bordawekar.
\newblock Evaluation of parallel hashing techniques.
\newblock In {\em GPU Technology Conference}, Mar. 2014.

\bibitem{Botelho:2007}
F.~C. Botelho and N.~Ziviani.
\newblock External perfect hashing for very large key sets.
\newblock In {\em Proceedings of the Sixteenth ACM Conference on Conference on
  Information and Knowledge Management}, CIKM '07, pp. 653--662. ACM, New York,
  NY, USA, 2007.

\bibitem{Brent:1974}
R.~P. Brent.
\newblock The parallel evaluation of general arithmetic expressions.
\newblock {\em J. ACM}, 21(2):201--206, Apr. 1974.

\bibitem{Breslow:2016}
\href{http://dl.acm.org/citation.cfm?id=3026959.3026986}{A.~D. Breslow, D.~P.
  Zhang, J.~L. Greathouse, N.~Jayasena, and D.~M. Tullsen}.
\newblock \href{http://dl.acm.org/citation.cfm?id=3026959.3026986}{Horton
  tables: Fast hash tables for in-memory data-intensive computing}.
\newblock \href{http://dl.acm.org/citation.cfm?id=3026959.3026986}{In {\em
  Proceedings of the 2016 USENIX Conference on Usenix Annual Technical
  Conference}}, \href{http://dl.acm.org/citation.cfm?id=3026959.3026986}{USENIX
  ATC '16}, \href{http://dl.acm.org/citation.cfm?id=3026959.3026986}{pp.
  281--294}. \href{http://dl.acm.org/citation.cfm?id=3026959.3026986}{USENIX
  Association},
  \href{http://dl.acm.org/citation.cfm?id=3026959.3026986}{Berkeley, CA, USA},
  \href{http://dl.acm.org/citation.cfm?id=3026959.3026986}{2016}.

\bibitem{Cederman:2012}
D.~Cederman, B.~Chatterjee, and P.~Tsigas.
\newblock Understanding the performance of concurrent data structures on
  graphics processors.
\newblock In {\em Proceedings of the 18th International Conference on European
  Parallel Processing}, Euro-Par 2012, pp. 883--894, August 2012.

\bibitem{Celis:1986}
P.~Celis.
\newblock {\em Robin Hood Hashing}.
\newblock PhD thesis, Waterloo, Ont., Canada, Canada, 1986.

\bibitem{Cheng:2014}
L.~Cheng, S.~Kotoulas, T.~E. Ward, and G.~Theodoropoulos.
\newblock Design and evaluation of parallel hashing over large-scale data.
\newblock In {\em 2014 21st International Conference on High Performance
  Computing (HiPC)}, pp. 1--10, Dec 2014.

\bibitem{Chentanez:2016}
\href{http://dx.doi.org/https://doi.org/10.1016/j.cag.2016.03.002}{N.~Chentanez,
  M.~M{\"u}ller, and M.~Macklin}.
\newblock
  \href{http://dx.doi.org/https://doi.org/10.1016/j.cag.2016.03.002}{{GPU}
  accelerated grid-free surface tracking}.
\newblock
  \href{http://dx.doi.org/https://doi.org/10.1016/j.cag.2016.03.002}{{\em
  Computers \& Graphics}},
  \href{http://dx.doi.org/https://doi.org/10.1016/j.cag.2016.03.002}{57(Supplement
  C):1 -- 11},
  \href{http://dx.doi.org/https://doi.org/10.1016/j.cag.2016.03.002}{2016}.
  \href{https://doi.org/10.1016/j.cag.2016.03.002}
{doi: \textsf{%
10\hspace{.1pt}\discretionary{.}{%
}{.}\hspace{.4pt}1016\discretionary{/}{%
}{/}j\hspace{.1pt}\discretionary{.}{%
}{.}\hspace{.4pt}cag\hspace{.1pt}\discretionary{.}{%
}{.}\hspace{.4pt}2016\hspace{.1pt}\discretionary{.}{%
}{.}\hspace{.4pt}03\hspace{.1pt}\discretionary{.}{%
}{.}\hspace{.4pt}002}}


\bibitem{Choi:2009}
M.~G. Choi, E.~Ju, J.-W. Chang, J.~Lee, and Y.~J. Kim.
\newblock Linkless octree using multi-level perfect hashing.
\newblock {\em Comput. Graph. Forum}, 28:1773--1780, 2009.

\bibitem{Choudhury:2016}
Z.~Choudhury, S.~Purini, and S.~R. Krishna.
\newblock A hybrid cpu+gpu working-set dictionary.
\newblock In {\em 2016 15th International Symposium on Parallel and Distributed
  Computing (ISPDC)}, pp. 56--63, July 2016.

\bibitem{Cormen:2001}
T.~H. Cormen, C.~Stein, R.~L. Rivest, and C.~E. Leiserson.
\newblock {\em Introduction to Algorithms}.
\newblock McGraw-Hill Higher Education, 2nd ed., 2001.

\bibitem{Czech:1997}
Z.~J. Czech, G.~Havas, and B.~S. Majewski.
\newblock Perfect hashing.
\newblock {\em Theoretical Computer Science}, 182(1):1 -- 143, 1997.

\bibitem{Dasgupta:2008}
S.~Dasgupta and Y.~Freund.
\newblock Random projection trees and low dimensional manifolds.
\newblock In {\em Proceedings of the Fortieth Annual ACM Symposium on Theory of
  Computing}, STOC '08, pp. 537--546. ACM, New York, NY, USA, 2008.

\bibitem{Datar:2004}
M.~Datar, N.~Immorlica, P.~Indyk, and V.~S. Mirrokni.
\newblock Locality-sensitive hashing scheme based on p-stable distributions.
\newblock In {\em Proceedings of the Twentieth Annual Symposium on
  Computational Geometry}, SCG '04, pp. 253--262. ACM, New York, NY, USA, 2004.

\bibitem{Davidson:2012}
A.~Davidson, D.~Tarjan, M.~Garland, and J.~D. Owens.
\newblock Efficient parallel merge sort for fixed and variable length keys.
\newblock In {\em Innovative Parallel Computing}, p.~9, May 2012.

\bibitem{Dice:2013}
D.~Dice, D.~Hendler, and I.~Mirsky.
\newblock Lightweight contention management for efficient compare-and-swap
  operations.
\newblock In {\em Proceedings of the 19th International Conference on Parallel
  Processing}, Euro-Par'13, pp. 595--606. Springer-Verlag, Berlin, Heidelberg,
  2013.

\bibitem{Duan:2017}
\href{http://dx.doi.org/https://doi.org/10.1016/j.cag.2017.08.012}{W.~Duan,
  J.~Luo, G.~Ni, B.~Tang, Q.~Hu, and Y.~Gao}.
\newblock
  \href{http://dx.doi.org/https://doi.org/10.1016/j.cag.2017.08.012}{Exclusive
  grouped spatial hashing}.
\newblock
  \href{http://dx.doi.org/https://doi.org/10.1016/j.cag.2017.08.012}{{\em
  Computers \& Graphics}},
  \href{http://dx.doi.org/https://doi.org/10.1016/j.cag.2017.08.012}{2017}.
  \href{https://doi.org/10.1016/j.cag.2017.08.012}
{doi: \textsf{%
10\hspace{.1pt}\discretionary{.}{%
}{.}\hspace{.4pt}1016\discretionary{/}{%
}{/}j\hspace{.1pt}\discretionary{.}{%
}{.}\hspace{.4pt}cag\hspace{.1pt}\discretionary{.}{%
}{.}\hspace{.4pt}2017\hspace{.1pt}\discretionary{.}{%
}{.}\hspace{.4pt}08\hspace{.1pt}\discretionary{.}{%
}{.}\hspace{.4pt}012}}


\bibitem{Eitz:2007}
\href{http://dx.doi.org/10.1109/SMI.2007.18}{M.~Eitz and G.~Lixu}.
\newblock \href{http://dx.doi.org/10.1109/SMI.2007.18}{Hierarchical spatial
  hashing for real-time collision detection}.
\newblock \href{http://dx.doi.org/10.1109/SMI.2007.18}{In {\em Shape Modeling
  and Applications, 2007. SMI '07. IEEE International Conference on}},
  \href{http://dx.doi.org/10.1109/SMI.2007.18}{pp. 61--70},
  \href{http://dx.doi.org/10.1109/SMI.2007.18}{June 2007}.
  \href{https://dx.doi.org/10.1109/SMI.2007.18}
{doi: \textsf{%
10\hspace{.1pt}\discretionary{.}{%
}{.}\hspace{.4pt}1109\discretionary{/}{%
}{/}SMI\hspace{.1pt}\discretionary{.}{%
}{.}\hspace{.4pt}2007\hspace{.1pt}\discretionary{.}{%
}{.}\hspace{.4pt}18}}


\bibitem{Erlingsson:2006}
Ã.~Erlingsson, M.~Manasse, and F.~McSherry.
\newblock A cool and practical alternative to traditional hash tables.
\newblock In {\em 7th Workshop on Distributed Data and Structures (WDAS'06)},
  pp. 1--6. Santa Clara, CA, January 2006.

\bibitem{Eyiyurekli:2011}
M.~Eyiyurekli and D.~E. Breen.
\newblock Data structures for interactive high resolution level-set surface
  editing.
\newblock In {\em Proceedings of Graphics Interface 2011}, GI '11, pp. 95--102.
  Canadian Human-Computer Communications Society, School of Computer Science,
  University of Waterloo, Waterloo, Ontario, Canada, 2011.

\bibitem{Flynn:1972}
M.~J. Flynn.
\newblock Some computer organizations and their effectiveness.
\newblock {\em IEEE Trans. Comput.}, 21(9):948--960, Sept. 1972.

\bibitem{Fox:1992}
E.~A. Fox, L.~S. Heath, Q.~F. Chen, and A.~M. Daoud.
\newblock Practical minimal perfect hash functions for large databases.
\newblock {\em Commun. ACM}, 35(1):105--121, Jan. 1992.

\bibitem{Fredman:1984}
M.~L. Fredman, J.~Koml\'{o}s, and E.~Szemer{\'e}di.
\newblock Storing a sparse table with 0(1) worst case access time.
\newblock {\em J. ACM}, 31(3):538--544, June 1984.

\bibitem{Gao:2015}
H.~Gao, J.~Tang, and G.~Wu.
\newblock Parallel surface reconstruction on gpu.
\newblock In {\em Proceedings of the 7th International Conference on Internet
  Multimedia Computing and Service}, ICIMCS '15, pp. 54:1--54:5. ACM, New York,
  NY, USA, 2015.

\bibitem{Garcia:2011}
I.~Garc\'{\i}a, S.~Lefebvre, S.~Hornus, and A.~Lasram.
\newblock Coherent parallel hashing.
\newblock {\em ACM Trans. Graph.}, 30(6):161:1--161:8, Dec. 2011.

\bibitem{Gieseke:2014}
F.~Gieseke, J.~Heinermann, C.~Oancea, and C.~Igel.
\newblock Buffer k-d trees: Processing massive nearest neighbor queries on
  {GPU}s.
\newblock 1:172--180, 01 2014.

\bibitem{Goodman:2010}
E.~L. Goodman, D.~J. Haglin, C.~Scherrer, D.~Chavarr\'{i}a-Miranda, J.~Mogill,
  and J.~Feo.
\newblock Hashing strategies for the cray xmt.
\newblock In {\em 2010 IEEE International Symposium on Parallel Distributed
  Processing, Workshops and Phd Forum (IPDPSW)}, pp. 1--8, April 2010.

\bibitem{Greenwald:2002}
M.~Greenwald.
\newblock Two-handed emulation: How to build non-blocking implementations of
  complex data-structures using dcas.
\newblock In {\em Proceedings of the Twenty-first Annual Symposium on
  Principles of Distributed Computing}, PODC '02, pp. 260--269. ACM, New York,
  NY, USA, 2002.

\bibitem{Gustafson:1988}
J.~L. Gustafson.
\newblock Reevaluating amdahl's law.
\newblock {\em Commun. ACM}, 31(5):532--533, May 1988.

\bibitem{Nvidia-Maxwell}
M.~Harris.
\newblock {Maxwell: The Most Advanced CUDA GPU Ever Made}.
\newblock
  \url{https://devblogs.nvidia.com/parallelforall/maxwell-most-advanced-cuda-gpu-ever-made/},
  2014.

\bibitem{He:2012}
X.~He, D.~Agarwal, and S.~K. Prasad.
\newblock Design and implementation of a parallel priority queue on many-core
  architectures.
\newblock In {\em 2012 19th International Conference on High Performance
  Computing}, pp. 1--10, Dec 2012.

\bibitem{Herlihy:1991}
M.~Herlihy.
\newblock Wait-free synchronization.
\newblock {\em ACM Trans. Program. Lang. Syst.}, 13(1):124--149, Jan. 1991.

\bibitem{Hetherington:2015}
T.~H. Hetherington, M.~O'Connor, and T.~M. Aamodt.
\newblock Memcachedgpu: Scaling-up scale-out key-value stores.
\newblock In {\em Proceedings of the Sixth ACM Symposium on Cloud Computing},
  SoCC '15, pp. 43--57. ACM, New York, NY, USA, 2015.

\bibitem{Huang:2010}
X.~Huang, C.~I. Rodrigues, S.~Jones, I.~Buck, and W.~m.~Hwu.
\newblock {XMalloc}: A scalable lock-free dynamic memory allocator for
  many-core machines.
\newblock In {\em 2010 10th IEEE International Conference on Computer and
  Information Technology}, pp. 1134--1139, June 2010.

\bibitem{Indyk:1998}
P.~Indyk and R.~Motwani.
\newblock Approximate nearest neighbors: Towards removing the curse of
  dimensionality.
\newblock In {\em Proceedings of the Thirtieth Annual ACM Symposium on Theory
  of Computing}, STOC '98, pp. 604--613. ACM, New York, NY, USA, 1998.

\bibitem{TBB:web:2017}
{Intel Corporation}.
\newblock {Introducing the Intel Threading Building Blocks}, May 2017.
\newblock \url{https://software.intel.com/en-us/node/506042}.

\bibitem{Jeffers:2015}
J.~Jeffers and J.~Reinders.
\newblock {\em High Performance Parallelism Pearls Volume Two: Multicore and
  Many-core Programming Approaches}, vol.~2.
\newblock Morgan Kaufmann Publishers Inc., San Francisco, CA, USA, 1st ed.,
  2015.

\bibitem{Kahler:2016}
\href{http://dx.doi.org/10.1109/LRA.2015.2512958}{O.~K{\"a}hler, V.~Prisacariu,
  J.~Valentin, and D.~Murray}.
\newblock \href{http://dx.doi.org/10.1109/LRA.2015.2512958}{Hierarchical voxel
  block hashing for efficient integration of depth images}.
\newblock \href{http://dx.doi.org/10.1109/LRA.2015.2512958}{{\em IEEE Robotics
  and Automation Letters}},
  \href{http://dx.doi.org/10.1109/LRA.2015.2512958}{1(1):192--197},
  \href{http://dx.doi.org/10.1109/LRA.2015.2512958}{Jan 2016}.
  \href{https://dx.doi.org/10.1109/LRA.2015.2512958}
{doi: \textsf{%
10\hspace{.1pt}\discretionary{.}{%
}{.}\hspace{.4pt}1109\discretionary{/}{%
}{/}LRA\hspace{.1pt}\discretionary{.}{%
}{.}\hspace{.4pt}2015\hspace{.1pt}\discretionary{.}{%
}{.}\hspace{.4pt}2512958}}


\bibitem{Kaldewey:2012}
T.~Kaldewey and A.~Di~Blas.
\newblock Large-scale gpu search.
\newblock pp. 3--14, 12 2012.

\bibitem{Kalojanov:2011}
J.~Kalojanov, M.~Billeter, and P.~Slusallek.
\newblock Two-level grids for ray tracing on gpus.
\newblock {\em Computer Graphics Forum}, 30(2):307--314, 2011.

\bibitem{Kalojanov:2009}
J.~Kalojanov and P.~Slusallek.
\newblock A parallel algorithm for construction of uniform grids.
\newblock In {\em Proceedings of the Conference on High Performance Graphics
  2009}, HPG '09, pp. 23--28. ACM, New York, NY, USA, 2009.

\bibitem{Kapasi:2003}
U.~J. Kapasi, S.~Rixner, W.~J. Dally, B.~Khailany, J.~H. Ahn, P.~Mattson, and
  J.~D. Owens.
\newblock Programmable stream processors.
\newblock {\em Computer}, 36(8):54--62, Aug. 2003.

\bibitem{Karlin:1986}
A.~R. Karlin and E.~Upfal.
\newblock Parallel hashing\—an efficient implementation of shared memory.
\newblock In {\em Proceedings of the Eighteenth Annual ACM Symposium on Theory
  of Computing}, STOC '86, pp. 160--168. ACM, New York, NY, USA, 1986.

\bibitem{Karnagel:2015}
T.~Karnagel, R.~Mueller, and G.~M. Lohman.
\newblock Optimizing gpu-accelerated group-by and aggregation.
\newblock In {\em ADMS@VLDB}, 2015.

\bibitem{Karras:2012}
T.~Karras.
\newblock {Maximizing Parallelism in the Construction of BVHs, Octrees, and k-d
  Trees}.
\newblock In C.~Dachsbacher, J.~Munkberg, and J.~Pantaleoni, eds., {\em
  Eurographics/ ACM SIGGRAPH Symposium on High Performance Graphics}, pp.
  33--37. The Eurographics Association, 2012.

\bibitem{Khorasani:2015}
F.~Khorasani, M.~E. Belviranli, R.~Gupta, and L.~N. Bhuyan.
\newblock Stadium hashing: Scalable and flexible hashing on gpus.
\newblock In {\em 2015 International Conference on Parallel Architecture and
  Compilation (PACT)}, pp. 63--74, Oct 2015.

\bibitem{Kim:2010}
C.~Kim, J.~Chhugani, N.~Satish, E.~Sedlar, A.~D. Nguyen, T.~Kaldewey, V.~W.
  Lee, S.~A. Brandt, and P.~Dubey.
\newblock Fast: Fast architecture sensitive tree search on modern cpus and
  gpus.
\newblock In {\em Proceedings of the 2010 ACM SIGMOD International Conference
  on Management of Data}, SIGMOD '10, pp. 339--350. ACM, New York, NY, USA,
  2010.

\bibitem{Knuth:1998}
D.~E. Knuth.
\newblock {\em The Art of Computer Programming, Volume 3: (2nd Ed.) Sorting and
  Searching}.
\newblock Addison Wesley Longman Publishing Co., Inc., Redwood City, CA, USA,
  1998.

\bibitem{Kuang:2009}
Q.~Kuang and L.~Zhao.
\newblock A practical {GPU} based {KNN} algorithm.
\newblock In {\em Proceedings of the Second Symposium on International Computer
  Science and Computational Technology (ISCSCT '09)}, pp. 151--155. Academy
  Publisher, Dec. 2009.

\bibitem{Lagae:2008}
A.~Lagae and P.~Dutr{\'e}.
\newblock Compact, fast and robust grids for ray tracing.
\newblock In {\em ACM SIGGRAPH 2008 Talks}, SIGGRAPH '08, pp. 20:1--20:1. ACM,
  New York, NY, USA, 2008.

\bibitem{Lamport:1978}
L.~Lamport.
\newblock Time, clocks, and the ordering of events in a distributed system.
\newblock {\em Commun. ACM}, 21(7):558--565, July 1978.

\bibitem{Lauterbach:2009}
C.~Lauterbach, M.~Garland, S.~Sengupta, D.~Luebke, and D.~Manocha.
\newblock Fast bvh construction on gpus.
\newblock {\em Computer Graphics Forum}, 28(2):375--384, 2009.

\bibitem{Lefebvre:2006}
S.~Lefebvre and H.~Hoppe.
\newblock Perfect spatial hashing.
\newblock In {\em ACM SIGGRAPH 2006 Papers}, SIGGRAPH '06, pp. 579--588. ACM,
  New York, NY, USA, 2006.

\bibitem{Lessley:EGPGV2016}
B.~Lessley, R.~Binyahib, R.~Maynard, and H.~Childs.
\newblock {External Facelist Calculation with Data-Parallel Primitives}.
\newblock In {\em Proceedings of EuroGraphics Symposium on Parallel Graphics
  and \linebreak Visualization (EGPGV)}, pp. 10--20. Groningen, The
  Netherlands, June 2016.

\bibitem{Lessley:LDAV17-1}
B.~Lessley, K.~Moreland, M.~Larsen, and H.~Childs.
\newblock {Techniques for Data-Parallel Searching for Duplicate Elements}.
\newblock In {\em Proceedings of IEEE Symposium on Large Data Analysis and
  Visualization (LDAV)}, pp. 1--5. Phoenix, AZ, Oct. 2017.

\bibitem{Li:2015}
S.~Li and N.~Amenta.
\newblock Brute-force k-nearest neighbors search on the gpu.
\newblock In {\em Proceedings of the 8th International Conference on Similarity
  Search and Applications - Volume 9371}, SISAP 2015, pp. 259--270.
  Springer-Verlag New York, Inc., New York, NY, USA, 2015.

\bibitem{Little:2011}
\href{http://dx.doi.org/10.1287/opre.1110.0940}{J.~D.~C. Little}.
\newblock \href{http://dx.doi.org/10.1287/opre.1110.0940}{Or forum---little's
  law as viewed on its 50th anniversary}.
\newblock \href{http://dx.doi.org/10.1287/opre.1110.0940}{{\em Oper. Res.}},
  \href{http://dx.doi.org/10.1287/opre.1110.0940}{59(3):536--549},
  \href{http://dx.doi.org/10.1287/opre.1110.0940}{May 2011}.
  \href{https://dx.doi.org/10.1287/opre.1110.0940}
{doi: \textsf{%
10\hspace{.1pt}\discretionary{.}{%
}{.}\hspace{.4pt}1287\discretionary{/}{%
}{/}opre\hspace{.1pt}\discretionary{.}{%
}{.}\hspace{.4pt}1110\hspace{.1pt}\discretionary{.}{%
}{.}\hspace{.4pt}0940}}


\bibitem{Lukac:2015}
\href{http://dx.doi.org/10.1007/978-981-287-134-3_14}{N.~Luka{\v{c}} and
  B.~{\v{Z}}alik}.
\newblock \href{http://dx.doi.org/10.1007/978-981-287-134-3_14}{{\em Fast
  Approximate k-Nearest Neighbours Search Using GPGPU}},
  \href{http://dx.doi.org/10.1007/978-981-287-134-3_14}{pp. 221--234}.
\newblock \href{http://dx.doi.org/10.1007/978-981-287-134-3_14}{Springer
  Singapore}, \href{http://dx.doi.org/10.1007/978-981-287-134-3_14}{Singapore},
  \href{http://dx.doi.org/10.1007/978-981-287-134-3_14}{2015}.
  \href{https://dx.doi.org/10.1007/978-981-287-134-3_14}
{doi: \textsf{%
10\hspace{.1pt}\discretionary{.}{%
}{.}\hspace{.4pt}1007\discretionary{/}{%
}{/}978\discretionary{%
}{-}{-}981\discretionary{%
}{-}{-}287\discretionary{%
}{-}{-}134\discretionary{%
}{-}{-}3\_14}}


\bibitem{Luo:2012}
\href{http://dx.doi.org/10.1109/ASPDAC.2012.6164973}{L.~Luo, M.~D.~F. Wong, and
  L.~Leong}.
\newblock \href{http://dx.doi.org/10.1109/ASPDAC.2012.6164973}{Parallel
  implementation of r-trees on the gpu}.
\newblock \href{http://dx.doi.org/10.1109/ASPDAC.2012.6164973}{In {\em 17th
  Asia and South Pacific Design Automation Conference}},
  \href{http://dx.doi.org/10.1109/ASPDAC.2012.6164973}{pp. 353--358},
  \href{http://dx.doi.org/10.1109/ASPDAC.2012.6164973}{Jan 2012}.
  \href{https://dx.doi.org/10.1109/ASPDAC.2012.6164973}
{doi: \textsf{%
10\hspace{.1pt}\discretionary{.}{%
}{.}\hspace{.4pt}1109\discretionary{/}{%
}{/}ASPDAC\hspace{.1pt}\discretionary{.}{%
}{.}\hspace{.4pt}2012\hspace{.1pt}\discretionary{.}{%
}{.}\hspace{.4pt}6164973}}


\bibitem{Lv:2007}
Q.~Lv, W.~Josephson, Z.~Wang, M.~Charikar, and K.~Li.
\newblock Multi-probe lsh: Efficient indexing for high-dimensional similarity
  search.
\newblock In {\em Proceedings of the 33rd International Conference on Very
  Large Data Bases}, VLDB '07, pp. 950--961. VLDB Endowment, 2007.

\bibitem{Maurer:1975}
W.~D. Maurer and T.~G. Lewis.
\newblock Hash table methods.
\newblock {\em ACM Comput. Surv.}, 7(1):5--19, Mar. 1975.

\bibitem{McCool:2012}
M.~McCool, J.~Reinders, and A.~Robison.
\newblock {\em Structured Parallel Programming: Patterns for Efficient
  Computation}.
\newblock Morgan Kaufmann Publishers Inc., San Francisco, CA, USA, 1st ed.,
  2012.

\bibitem{Mehlhorn:1982}
\href{http://dx.doi.org/10.1109/SFCS.1982.80}{K.~Mehlhorn}.
\newblock \href{http://dx.doi.org/10.1109/SFCS.1982.80}{On the program size of
  perfect and universal hash functions}.
\newblock \href{http://dx.doi.org/10.1109/SFCS.1982.80}{In {\em 23rd Annual
  Symposium on Foundations of Computer Science (sfcs 1982)}},
  \href{http://dx.doi.org/10.1109/SFCS.1982.80}{pp. 170--175},
  \href{http://dx.doi.org/10.1109/SFCS.1982.80}{Nov 1982}.
  \href{https://dx.doi.org/10.1109/SFCS.1982.80}
{doi: \textsf{%
10\hspace{.1pt}\discretionary{.}{%
}{.}\hspace{.4pt}1109\discretionary{/}{%
}{/}SFCS\hspace{.1pt}\discretionary{.}{%
}{.}\hspace{.4pt}1982\hspace{.1pt}\discretionary{.}{%
}{.}\hspace{.4pt}80}}


\bibitem{Merrill:2010}
\href{http://dx.doi.org/10.1145/1854273.1854344}{D.~G. Merrill and A.~S.
  Grimshaw}.
\newblock \href{http://dx.doi.org/10.1145/1854273.1854344}{Revisiting sorting
  for gpgpu stream architectures}.
\newblock \href{http://dx.doi.org/10.1145/1854273.1854344}{In {\em Proceedings
  of the 19th International Conference on Parallel Architectures and
  Compilation Techniques}},
  \href{http://dx.doi.org/10.1145/1854273.1854344}{PACT '10},
  \href{http://dx.doi.org/10.1145/1854273.1854344}{pp. 545--546}.
  \href{http://dx.doi.org/10.1145/1854273.1854344}{ACM},
  \href{http://dx.doi.org/10.1145/1854273.1854344}{New York, NY, USA},
  \href{http://dx.doi.org/10.1145/1854273.1854344}{2010}.
  \href{https://dx.doi.org/10.1145/1854273.1854344}
{doi: \textsf{%
10\hspace{.1pt}\discretionary{.}{%
}{.}\hspace{.4pt}1145\discretionary{/}{%
}{/}1854273\hspace{.1pt}\discretionary{.}{%
}{.}\hspace{.4pt}1854344}}


\bibitem{Michael:2002}
\href{http://dx.doi.org/10.1145/564870.564881}{M.~M. Michael}.
\newblock \href{http://dx.doi.org/10.1145/564870.564881}{High performance
  dynamic lock-free hash tables and list-based sets}.
\newblock \href{http://dx.doi.org/10.1145/564870.564881}{In {\em Proceedings of
  the Fourteenth Annual ACM Symposium on Parallel Algorithms and
  Architectures}}, \href{http://dx.doi.org/10.1145/564870.564881}{SPAA '02},
  \href{http://dx.doi.org/10.1145/564870.564881}{pp. 73--82}.
  \href{http://dx.doi.org/10.1145/564870.564881}{ACM},
  \href{http://dx.doi.org/10.1145/564870.564881}{New York, NY, USA},
  \href{http://dx.doi.org/10.1145/564870.564881}{2002}.
  \href{https://dx.doi.org/10.1145/564870.564881}
{doi: \textsf{%
10\hspace{.1pt}\discretionary{.}{%
}{.}\hspace{.4pt}1145\discretionary{/}{%
}{/}564870\hspace{.1pt}\discretionary{.}{%
}{.}\hspace{.4pt}564881}}


\bibitem{Misra:2012}
P.~Misra and M.~Chaudhuri.
\newblock Performance evaluation of concurrent lock-free data structures on
  gpus.
\newblock In {\em Proceedings of the 2012 IEEE 18th International Conference on
  Parallel and Distributed Systems}, ICPADS '12, pp. 53--60. IEEE Computer
  Society, Washington, DC, USA, 2012.

\bibitem{Moazeni:2012}
M.~Moazeni and M.~Sarrafzadeh.
\newblock Lock-free hash table on graphics processors.
\newblock In {\em 2012 Symposium on Application Accelerators in High
  Performance Computing}, pp. 133--136, July 2012.

\bibitem{Moreland:CGA2016}
K.~Moreland, C.~Sewell, W.~Usher, L.~Lo, J.~Meredith, D.~Pugmire, J.~Kress,
  H.~Schroots, K.-L. Ma, H.~Childs, M.~Larsen, C.-M. Chen, R.~Maynard, and
  B.~Geveci.
\newblock {VTK-m: Accelerating the Visualization Toolkit for Massively Threaded
  Architectures}.
\newblock {\em IEEE Computer Graphics and Applications (CG\&A)}, 36(3):48--58,
  May/June 2016.

\bibitem{Moscovici:2017}
N.~Moscovici, N.~Cohen, and E.~Petrank.
\newblock A {GPU}-friendly skiplist algorithm.
\newblock In {\em 2017 26th International Conference on Parallel Architectures
  and Compilation Techniques (PACT)}, pp. 246--259, Sept 2017.

\bibitem{Munro:1992}
J.~I. Munro, T.~Papadakis, and R.~Sedgewick.
\newblock Deterministic skip lists.
\newblock In {\em Proceedings of the Third Annual ACM-SIAM Symposium on
  Discrete Algorithms}, SODA '92, pp. 367--375. Society for Industrial and
  Applied Mathematics, Philadelphia, PA, USA, 1992.

\bibitem{Niener:2013}
M.~Nie{\ss}ner, M.~Zollh\"ofer, S.~Izadi, and M.~Stamminger.
\newblock Real-time 3d reconstruction at scale using voxel hashing.
\newblock {\em ACM Transactions on Graphics (TOG)}, 2013.

\bibitem{Nvidia-BestPractices}
{Nvidia Corporation}.
\newblock {CUDA C Best Practices Guide}.
\newblock
  \url{http://docs.nvidia.com/cuda/cuda-c-best-practices-guide/index.html},
  2017.

\bibitem{Nvidia-ProgramGuide}
{Nvidia Corporation}.
\newblock {CUDA C Programming Guide}.
\newblock
  \url{http://docs.nvidia.com/cuda/cuda-c-programming-guide/index.html}, 2017.

\bibitem{Nvidia-PTX}
{Nvidia Corporation}.
\newblock {Parallel Thread Execution ISA Version 6.0}.
\newblock
  \url{http://docs.nvidia.com/cuda/parallel-thread-execution/index.html}, 2017.

\bibitem{Thrust:web:2017}
{Nvidia Corporation}.
\newblock {Thrust}, Nov. 2017.
\newblock \url{http://thrust.github.io}.

\bibitem{Owens:2007:Survey}
\href{http://dx.doi.org/10.1111/j.1467-8659.2007.01012.x}{J.~D. Owens,
  D.~Luebke, N.~Govindaraju, M.~Harris, J.~Krüger, A.~E. Lefohn, and T.~J.
  Purcell}.
\newblock \href{http://dx.doi.org/10.1111/j.1467-8659.2007.01012.x}{A survey of
  general-purpose computation on graphics hardware}.
\newblock \href{http://dx.doi.org/10.1111/j.1467-8659.2007.01012.x}{{\em
  Computer Graphics Forum}},
  \href{http://dx.doi.org/10.1111/j.1467-8659.2007.01012.x}{26(1):80--113},
  \href{http://dx.doi.org/10.1111/j.1467-8659.2007.01012.x}{2007}.
  \href{https://dx.doi.org/10.1111/j.1467-8659.2007.01012.x}
{doi: \textsf{%
10\hspace{.1pt}\discretionary{.}{%
}{.}\hspace{.4pt}1111\discretionary{/}{%
}{/}j\hspace{.1pt}\discretionary{.}{%
}{.}\hspace{.4pt}1467\discretionary{%
}{-}{-}8659\hspace{.1pt}\discretionary{.}{%
}{.}\hspace{.4pt}2007\hspace{.1pt}\discretionary{.}{%
}{.}\hspace{.4pt}01012\hspace{.1pt}\discretionary{.}{%
}{.}\hspace{.4pt}x}}


\bibitem{Pagh:2004}
R.~Pagh and F.~F. Rodler.
\newblock Cuckoo hashing.
\newblock {\em J. Algorithms}, 51(2):122--144, May 2004.

\bibitem{Pan:2010}
\href{http://dx.doi.org/10.1109/IROS.2010.5651449}{J.~Pan, C.~Lauterbach, and
  D.~Manocha}.
\newblock \href{http://dx.doi.org/10.1109/IROS.2010.5651449}{Efficient
  nearest-neighbor computation for gpu-based motion planning}.
\newblock \href{http://dx.doi.org/10.1109/IROS.2010.5651449}{In {\em 2010
  IEEE/RSJ International Conference on Intelligent Robots and Systems}},
  \href{http://dx.doi.org/10.1109/IROS.2010.5651449}{pp. 2243--2248},
  \href{http://dx.doi.org/10.1109/IROS.2010.5651449}{Oct 2010}.
  \href{https://dx.doi.org/10.1109/IROS.2010.5651449}
{doi: \textsf{%
10\hspace{.1pt}\discretionary{.}{%
}{.}\hspace{.4pt}1109\discretionary{/}{%
}{/}IROS\hspace{.1pt}\discretionary{.}{%
}{.}\hspace{.4pt}2010\hspace{.1pt}\discretionary{.}{%
}{.}\hspace{.4pt}5651449}}


\bibitem{Pan:2011}
J.~Pan and D.~Manocha.
\newblock Fast gpu-based locality sensitive hashing for k-nearest neighbor
  computation.
\newblock In {\em Proceedings of the 19th ACM SIGSPATIAL International
  Conference on Advances in Geographic Information Systems}, GIS '11, pp.
  211--220. ACM, New York, NY, USA, 2011.

\bibitem{Patterson:2008}
D.~A. Patterson and J.~L. Hennessy.
\newblock {\em Computer Organization and Design, Fourth Edition, Fourth
  Edition: The Hardware/Software Interface (The Morgan Kaufmann Series in
  Computer Architecture and Design)}.
\newblock Morgan Kaufmann Publishers Inc., San Francisco, CA, USA, 4th ed.,
  2008.

\bibitem{Peierls:2005}
T.~Peierls, B.~Goetz, J.~Bloch, J.~Bowbeer, D.~Lea, and D.~Holmes.
\newblock {\em Java Concurrency in Practice}.
\newblock Addison-Wesley Professional, 2005.

\bibitem{Plauger:2000}
P.~Plauger, M.~Lee, D.~Musser, and A.~A. Stepanov.
\newblock {\em C++ Standard Template Library}.
\newblock Prentice Hall PTR, Upper Saddle River, NJ, USA, 1st ed., 2000.

\bibitem{Polychroniou:2013}
O.~Polychroniou and K.~A. Ross.
\newblock High throughput heavy hitter aggregation for modern simd processors.
\newblock In {\em Proceedings of the Ninth International Workshop on Data
  Management on New Hardware}, DaMoN '13, pp. 6:1--6:6. ACM, New York, NY, USA,
  2013.

\bibitem{Pouchol:2009}
\href{http://dx.doi.org/10.1080/2151237X.2009.10129281}{M.~Pouchol, A.~Ahmad,
  B.~Crespin, and O.~Terraz}.
\newblock \href{http://dx.doi.org/10.1080/2151237X.2009.10129281}{A
  hierarchical hashing scheme for nearest neighbor search and broad-phase
  collision detection}.
\newblock \href{http://dx.doi.org/10.1080/2151237X.2009.10129281}{{\em Journal
  of Graphics, GPU, and Game Tools}},
  \href{http://dx.doi.org/10.1080/2151237X.2009.10129281}{14(2):45--59},
  \href{http://dx.doi.org/10.1080/2151237X.2009.10129281}{2009}.
  \href{https://dx.doi.org/10.1080/2151237X.2009.10129281}
{doi: \textsf{%
10\hspace{.1pt}\discretionary{.}{%
}{.}\hspace{.4pt}1080\discretionary{/}{%
}{/}2151237X\hspace{.1pt}\discretionary{.}{%
}{.}\hspace{.4pt}2009\hspace{.1pt}\discretionary{.}{%
}{.}\hspace{.4pt}10129281}}


\bibitem{Satish:2009}
\href{http://dx.doi.org/10.1109/IPDPS.2009.5161005}{N.~Satish, M.~Harris, and
  M.~Garland}.
\newblock \href{http://dx.doi.org/10.1109/IPDPS.2009.5161005}{Designing
  efficient sorting algorithms for manycore gpus}.
\newblock \href{http://dx.doi.org/10.1109/IPDPS.2009.5161005}{In {\em
  Proceedings of the 2009 IEEE International Symposium on Parallel\&Distributed
  Processing}}, \href{http://dx.doi.org/10.1109/IPDPS.2009.5161005}{IPDPS '09},
  \href{http://dx.doi.org/10.1109/IPDPS.2009.5161005}{pp. 1--10}.
  \href{http://dx.doi.org/10.1109/IPDPS.2009.5161005}{IEEE Computer Society},
  \href{http://dx.doi.org/10.1109/IPDPS.2009.5161005}{Washington, DC, USA},
  \href{http://dx.doi.org/10.1109/IPDPS.2009.5161005}{2009}.
  \href{https://dx.doi.org/10.1109/IPDPS.2009.5161005}
{doi: \textsf{%
10\hspace{.1pt}\discretionary{.}{%
}{.}\hspace{.4pt}1109\discretionary{/}{%
}{/}IPDPS\hspace{.1pt}\discretionary{.}{%
}{.}\hspace{.4pt}2009\hspace{.1pt}\discretionary{.}{%
}{.}\hspace{.4pt}5161005}}


\bibitem{Schlegel:2009}
B.~Schlegel, R.~Gemulla, and W.~Lehner.
\newblock K-ary search on modern processors.
\newblock In {\em Proceedings of the Fifth International Workshop on Data
  Management on New Hardware}, DaMoN '09, pp. 52--60. ACM, New York, NY, USA,
  2009.

\bibitem{Schneider:2017}
J.~Schneider and P.~Rautek.
\newblock A versatile and efficient gpu data structure for spatial indexing.
\newblock {\em IEEE Transactions on Visualization and Computer Graphics},
  23(1):911--920, Jan 2017.

\bibitem{VTK:Book}
W.~J. Schroeder, B.~Lorensen, and K.~Martin.
\newblock {\em {The Visualization Toolkit: An object-oriented approach to 3D
  graphics}}.
\newblock Kitware, 2004.

\bibitem{Scogland:2015}
T.~R. Scogland and W.-c. Feng.
\newblock Design and evaluation of scalable concurrent queues for many-core
  architectures.
\newblock In {\em Proceedings of the 6th ACM/SPEC International Conference on
  Performance Engineering}, ICPE '15, pp. 63--74. ACM, New York, NY, USA, 2015.

\bibitem{Shalev:2006}
O.~Shalev and N.~Shavit.
\newblock Split-ordered lists: Lock-free extensible hash tables.
\newblock {\em J. ACM}, 53(3):379--405, May 2006.

\bibitem{Singh:2017}
D.~P. Singh, I.~Joshi, and J.~Choudhary.
\newblock Survey of gpu based sorting algorithms.
\newblock {\em International Journal of Parallel Programming}, Apr 2017.

\bibitem{Steinberger:2012}
M.~Steinberger, M.~Kenzel, B.~Kainz, and D.~Schmalstieg.
\newblock {ScatterAlloc}: Massively parallel dynamic memory allocation for the
  {GPU}.
\newblock In {\em 2012 Innovative Parallel Computing (InPar)}, pp. 1--10, May
  2012.

\bibitem{Stuart:2011}
J.~A. Stuart and J.~D. Owens.
\newblock Efficient synchronization primitives for {GPU}s.
\newblock {\em CoRR}, abs/1110.4623(1110.4623v1), Oct. 2011.

\bibitem{Suzuki:2006}
K.~Suzuki, D.~Tonien, K.~Kurosawa, and K.~Toyota.
\newblock {\em Birthday Paradox for Multi-collisions}, pp. 29--40.
\newblock Springer Berlin Heidelberg, Berlin, Heidelberg, 2006.

\bibitem{Teschner:2003}
M.~Teschner, B.~Heidelberger, M.~Mueller, D.~Pomeranets, and M.~Gross.
\newblock Optimized spatial hashing for collision detection of deformable
  objects.
\newblock {\em Proceedings of Vision, Modeling, Visualization (VMV 2003)}, pp.
  47--54, 2003.

\bibitem{Todd:2016}
A.~Todd, H.~Truong, J.~Deters, J.~Long, G.~Conant, and M.~Becchi.
\newblock Parallel gene upstream comparison via multi-level hash tables on gpu.
\newblock In {\em 2016 IEEE 22nd International Conference on Parallel and
  Distributed Systems (ICPADS)}, pp. 1049--1058, Dec 2016.

\bibitem{Tumblin:2015}
\href{http://dx.doi.org/10.1137/13093371X}{R.~Tumblin, P.~Ahrens, S.~Hartse,
  and R.~W. Robey}.
\newblock \href{http://dx.doi.org/10.1137/13093371X}{Parallel compact hash
  algorithms for computational meshes}.
\newblock \href{http://dx.doi.org/10.1137/13093371X}{{\em SIAM Journal on
  Scientific Computing}},
  \href{http://dx.doi.org/10.1137/13093371X}{37(1):C31--C53},
  \href{http://dx.doi.org/10.1137/13093371X}{2015}.
  \href{https://dx.doi.org/10.1137/13093371X}
{doi: \textsf{%
10\hspace{.1pt}\discretionary{.}{%
}{.}\hspace{.4pt}1137\discretionary{/}{%
}{/}13093371X}}


\bibitem{Ullman:1972}
J.~D. Ullman.
\newblock A note on the efficiency of hashing functions.
\newblock {\em J. ACM}, 19(3):569--575, July 1972.

\bibitem{Vinkler:2015}
M.~Vinkler and V.~Havran.
\newblock Register efficient dynamic memory allocator for gpus.
\newblock {\em Computer Graphics Forum}, (8):143--154, 2015.

\bibitem{Volkov:2016}
\href{http://www2.eecs.berkeley.edu/Pubs/TechRpts/2016/EECS-2016-143.html}{V.~Volkov}.
\newblock
  \href{http://www2.eecs.berkeley.edu/Pubs/TechRpts/2016/EECS-2016-143.html}{{\em
  Understanding Latency Hiding on GPUs}}.
\newblock
  \href{http://www2.eecs.berkeley.edu/Pubs/TechRpts/2016/EECS-2016-143.html}{PhD
  thesis},
  \href{http://www2.eecs.berkeley.edu/Pubs/TechRpts/2016/EECS-2016-143.html}{EECS
  Department, University of California, Berkeley},
  \href{http://www2.eecs.berkeley.edu/Pubs/TechRpts/2016/EECS-2016-143.html}{Aug
  2016}.

\bibitem{Wang:2016}
\href{http://dx.doi.org/10.1109/JPROC.2015.2487976}{J.~Wang, W.~Liu, S.~Kumar,
  and S.~F. Chang}.
\newblock \href{http://dx.doi.org/10.1109/JPROC.2015.2487976}{Learning to hash
  for indexing big data---{A} survey}.
\newblock \href{http://dx.doi.org/10.1109/JPROC.2015.2487976}{{\em Proceedings
  of the IEEE}},
  \href{http://dx.doi.org/10.1109/JPROC.2015.2487976}{104(1):34--57},
  \href{http://dx.doi.org/10.1109/JPROC.2015.2487976}{Jan 2016}.
  \href{https://dx.doi.org/10.1109/JPROC.2015.2487976}
{doi: \textsf{%
10\hspace{.1pt}\discretionary{.}{%
}{.}\hspace{.4pt}1109\discretionary{/}{%
}{/}JPROC\hspace{.1pt}\discretionary{.}{%
}{.}\hspace{.4pt}2015\hspace{.1pt}\discretionary{.}{%
}{.}\hspace{.4pt}2487976}}


\bibitem{Wang:2014}
\href{http://arxiv.org/abs/1408.2927}{J.~Wang, H.~T. Shen, J.~Song, and J.~Ji}.
\newblock \href{http://arxiv.org/abs/1408.2927}{Hashing for similarity search:
  {A} survey}.
\newblock \href{http://arxiv.org/abs/1408.2927}{{\em CoRR}},
  \href{http://arxiv.org/abs/1408.2927}{abs/1408.2927},
  \href{http://arxiv.org/abs/1408.2927}{2014}.

\bibitem{Widanagamaachchi:2014}
W.~Widanagamaachchi, P.~T. Bremer, C.~Sewell, L.~T. Lo, J.~Ahrens, and
  V.~Pascuccik.
\newblock Data-parallel halo finding with variable linking lengths.
\newblock In {\em 2014 IEEE 4th Symposium on Large Data Analysis and
  Visualization (LDAV)}, pp. 27--34, Nov 2014.

\bibitem{Yang:2010}
J.~C. Yang, J.~Hensley, H.~Gr\"{u}n, and N.~Thibieroz.
\newblock Real-time concurrent linked list construction on the gpu.
\newblock In {\em Proceedings of the 21st Eurographics Conference on
  Rendering}, EGSR'10, pp. 1297--1304. Eurographics Association, Aire-la-Ville,
  Switzerland, Switzerland, 2010.

\bibitem{Zhang:2015}
\href{http://dx.doi.org/10.14778/2809974.2809984}{K.~Zhang, K.~Wang, Y.~Yuan,
  L.~Guo, R.~Lee, and X.~Zhang}.
\newblock \href{http://dx.doi.org/10.14778/2809974.2809984}{Mega-kv: A case for
  gpus to maximize the throughput of in-memory key-value stores}.
\newblock \href{http://dx.doi.org/10.14778/2809974.2809984}{{\em Proc. VLDB
  Endow.}},
  \href{http://dx.doi.org/10.14778/2809974.2809984}{8(11):1226--1237},
  \href{http://dx.doi.org/10.14778/2809974.2809984}{July 2015}.
  \href{https://dx.doi.org/10.14778/2809974.2809984}
{doi: \textsf{%
10\hspace{.1pt}\discretionary{.}{%
}{.}\hspace{.4pt}14778\discretionary{/}{%
}{/}2809974\hspace{.1pt}\discretionary{.}{%
}{.}\hspace{.4pt}2809984}}


\bibitem{Zhou:2016}
\href{http://arxiv.org/abs/1603.08390}{J.~Zhou, Q.~Guo, H.~V. Jagadish,
  W.~Luan, A.~K.~H. Tung, Y.~Yang, and Y.~Zheng}.
\newblock \href{http://arxiv.org/abs/1603.08390}{Generic inverted index on the
  {GPU}}.
\newblock \href{http://arxiv.org/abs/1603.08390}{{\em CoRR}},
  \href{http://arxiv.org/abs/1603.08390}{abs/1603.08390},
  \href{http://arxiv.org/abs/1603.08390}{2016}.

\bibitem{Zhou:2011}
K.~Zhou, M.~Gong, X.~Huang, and B.~Guo.
\newblock Data-parallel octrees for surface reconstruction.
\newblock {\em IEEE Transactions on Visualization and Computer Graphics},
  17(5):669--681, May 2011.

\bibitem{Zhou:2008}
K.~Zhou, Q.~Hou, R.~Wang, and B.~Guo.
\newblock Real-time kd-tree construction on graphics hardware.
\newblock In {\em ACM SIGGRAPH Asia 2008 Papers}, SIGGRAPH Asia '08, pp.
  126:1--126:11. ACM, New York, NY, USA, 2008.

\end{thebibliography}

\end{document}